\documentclass[iop,numberedappendix]{emulateapj}
\newcommand{\lacc}{$L_{\rm acc}$}
\usepackage{natbib}
\usepackage{enumerate}
\usepackage{color}
\usepackage{threeparttable}
\usepackage{tablefootnote}
\usepackage{subfigure}
\usepackage{graphicx}
\usepackage{amssymb}

\begin{document}

\title{Indirect Detection of Forming Protoplanets via Chemical Asymmetries in Disks}
\shorttitle{Embedded Planet Detection via Chemical Asymmetries}
   
 \author{L. Ilsedore Cleeves\altaffilmark{1}, Edwin A. Bergin\altaffilmark{1}, Tim J. Harries\altaffilmark{2}}
\shortauthors{Cleeves, Bergin, and Harries}
\altaffiltext{1}{Department of Astronomy, University of Michigan, 1085 S. University Ave, Ann Arbor, MI 48109}
\altaffiltext{2}{Department of Physics and Astronomy, University of Exeter, Stocker Road, Exeter EX4 4QL}

\begin{abstract}
We examine changes in the molecular abundances resulting from increased heating due to a self-luminous planetary companion embedded within a narrow circumstellar disk gap.  Using 3D models that include stellar and planetary irradiation, we find that luminous young planets locally heat up the parent circumstellar disk by many tens of Kelvin, resulting in efficient thermal desorption of molecular species that are otherwise locally frozen out.  Furthermore, the heating is deposited over large regions of the disk, $\pm5$~AU radially and spanning $\lesssim60^\circ$ azimuthally.  From the 3D chemical models, we compute rotational line emission models and full ALMA simulations, and find that the chemical signatures of the young planet are detectable as chemical asymmetries in $\sim10h$ observations. HCN and its isotopologues are particularly clear tracers of planetary heating for the models considered here, and emission from multiple transitions of the same species is detectable, which encodes temperature information in addition to possible velocity information from the spectra itself.  We find submillimeter molecular emission will be a useful tool to study gas giant planet formation in situ, especially beyond $R\gtrsim10$~AU.  
\end{abstract}


\section{Introduction} 
Planetary systems form from the accretion disks encircling young stars.  The composition of ice, gas and dust within the disk sets the initial chemical conditions of the planets and also regulates the physical conditions under which planets form.  The properties of the pre-planetary materials can then be compared to the composition of the present day solar system \citep[e.g.][]{oberg2011co} and even that of extrasolar planetary atmospheres \citep{madhu2012,teske2013}.
Nonetheless, there is a missing link between these two stages, separated in time by billions of years.  It is essential to observationally capture a forming young planet in situ to put together a complete chemical (and physical) history of planet formation.  

The upcoming capabilities of the Atacama Large Millimeter Array (ALMA) will provide extremely high sensitivity and spatially resolved observations of disks (up to 0.007$''$ at 650 GHz, or 0.7 AU at distances of $d=100$~pc).  At these scales, detailed disk structure will be readily revealed, and observations of the local environment near forming protoplanets should be accessible.  Young proto-Jupiters are expected to be intrinsically hot as they accrete matter through their circumplanetary disk and liberate gravitational potential energy, thereby generating substantial accretion luminosity, \lacc.  Theoretical models of early-stage circumplanetary disks find typical ``quiescent'' accretion levels between $\dot{M}=10^{-10}$~M$_\odot$~year$^{-1}$ and $10^{-8}$ ~M$_\odot$~year$^{-1}$ \citep{ayliffe2009,lubow2012}.  Periodically, the circumplanetary disk is theorized to undergo accretion outbursts similar to those seen in FU Ori objects \citep{hartmann1985,zhu2009}, where the planet's accretion rate jumps to $\dot{M}=(1-10)\times10^{-5}$~M$_\odot$~year$^{-1}$.  Accretion rates can be translated into accretion luminosities, 
\begin{equation}
L_{\rm acc} = \frac{GM_p\dot{M_p}}{2R_{\rm in}}
\end{equation}
\citep{pringle1981}, where $M_p$ and $\dot{M_p}$ are the planet's mass and accretion rate, and $R_{\rm in}$ is the inner radius of the circumplanetary disk, which is something like the radius of the planet.  Assuming a Jupiter-mass planet and $R_{\rm in}=3$~R$_{\rm Jup}$, the quiescent accretion levels translate to $L_{\rm acc}=(3-300)\times10^{-5}$~L$_\odot$ \citep[see also][]{montesinos2015}.  During an accretion burst, however, the planet can outshine even the star in bolometric luminosity, where the accretion luminosity due to the planet can be as high as $L_{\rm acc}\sim15$~L$_\odot$ but for less than 0.1\% of the planet's formation time \citep{lubow2012}.   \citet{kraus2012} reported the detection of a possible embedded protoplanet companion in LkCa~15 for which they estimated a substantial accretion luminosity of \lacc$ = 10^{-3}$~L$_\odot$.   Planets with similar accretion luminosities will heat the nearby circumstellar dust disk and may give rise to detectable signatures with sensitive submillimeter continuum observations, though detecting the circumplanetary disk itself will be challenging \citep{wolf2005}.

In addition to the thermal effects on the dust, the planetary accretion heating will have a substantial impact on the chemical structure in the vicinity of the young planet.  In the present work, we explore local heating due to a single massive protoplanet (a gas giant precursor) on the three-dimensional chemistry of the surrounding circumstellar disk.  We do not calculate the chemistry of the hot, young circumplanetary disk, but focus instead on the larger scale effects on the cold, molecular disk within which the planet is entrenched.   We vary the planet's orbital location from the star for a fixed accretion rate of $\dot{M}=10^{-8}$~M$_\odot$~year$^{-1}$, where the physical and thermal structure are described in \S\ref{sec:phys}.  On this model, we calculate the time-dependent chemistry in 2D slices over the 3D model, focusing the calculations on the region near the planet and opposite the planet (\S\ref{sec:chem}).   For species that are particularly sensitive to the presence of the planet and have strong submillimeter transitions, we identify submillimeter emission line tracers and compute the emission based on the three-dimensional structure to determine detectability (\S\ref{sec:emission}). In \S\ref{sec:considerations} we relax some of the assumptions of our model and discuss potential caveats to our simple approach, and \S\ref{sec:conclusions}  summarizes our findings.

\section{Physical Model}\label{sec:phys}
\subsection{Axisymmetric Structure}\label{sec:den}
\subsubsection{Density Model}
\begin{figure}[ht]
\begin{centering}
\includegraphics[width=0.4\textwidth]{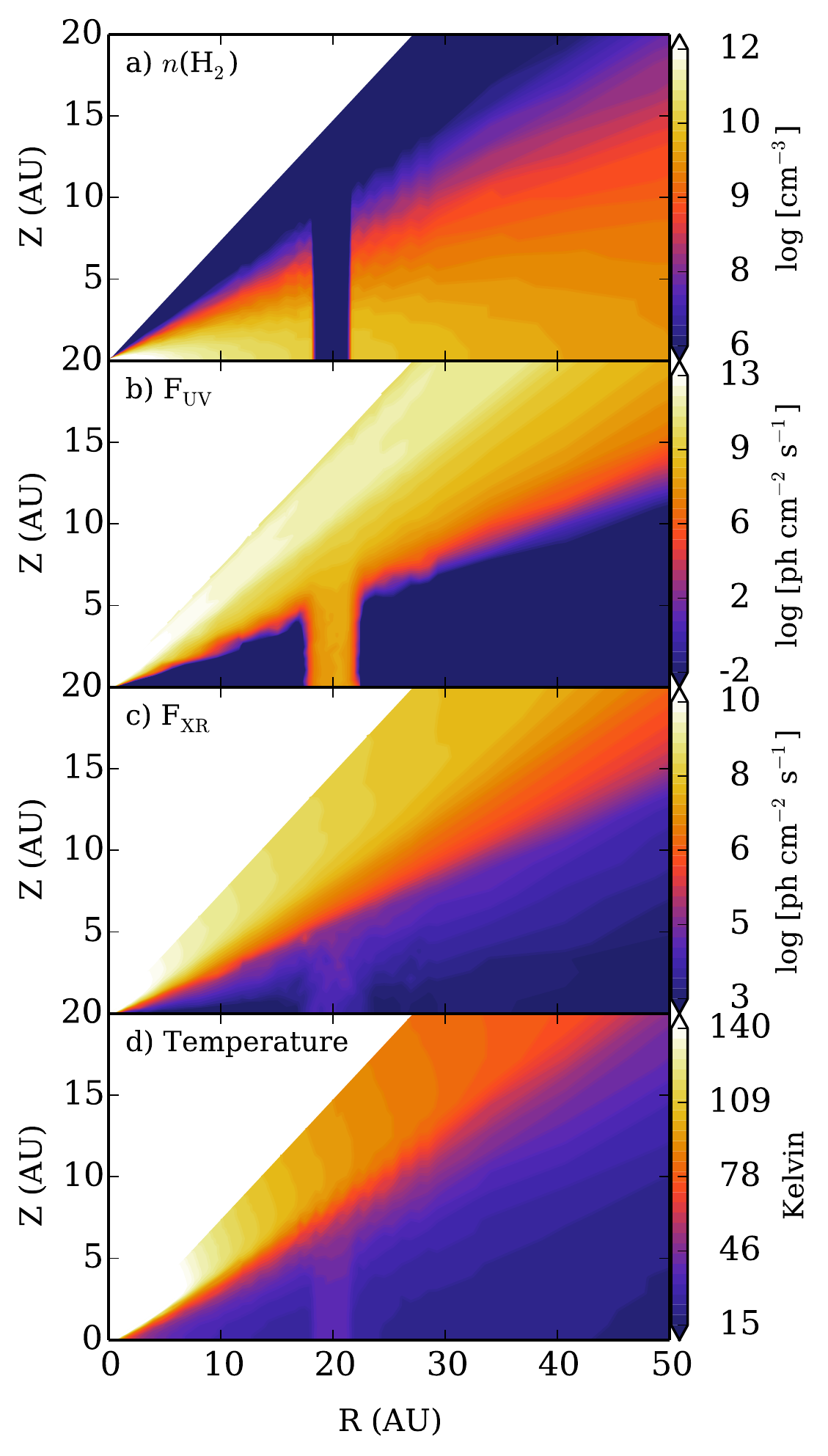}
\caption{Circumstellar disk properties shown for the planet at $d_p=20$~AU.  Disk properties away from the gap are similar for the $d_p=5$~AU and $d_p=10$~AU. The central star is located at R=0 and Z=0 and the plot is made for an azimuthal angle of $\phi=270^\circ$ (opposite the location of planet).\label{fig:fullmod}}
\end{centering}
\end{figure}

The background disk density is described by a simple power law in surface density, $\Sigma_g\propto R^{-1}$ and vertically gaussian with a scale height of $h\propto 12.5 \left (r/100~{\rm AU} \right )^\beta$~AU, $\beta=1.1$.  The gas and dust are uniformly mixed with a gas-to-dust mass ratio of 100, and the disk gas mass contains $M_g=0.01$~M$_{\odot}$. The dust is a simple MRN distribution in grain size \citep[$a_{\rm gr}\propto r_{\rm gr}^{-3.5}$;][]{mrn1977} with minimum size of 0.005~$\mu$m and maximum size of 1~$\mu$m.  The dust composition is assumed to be Draine and Lee astrosilicates \citep{draine1984}. We fix the central star to have a mass of $M_*=0.5$~M$_{\odot}$, an effective temperature of $T_{\rm eff} = 4000$~K and $R_*=2.0$~R$_{\odot}$, characteristic of low mass T Tauri stars.  We consider four planet locations, $d_p=5$~AU, 10~AU, 20~AU, and 30~AU. For each location, we simply define a vertical gap in the disk corresponding to the Hill radius of a $M_p=1$~M$_{\rm Jup}$ planet as consistent with theoretical models, which find massive planets open radial gaps comparable to the size of their Hill sphere \citep[e.g.,][]{lin1986,lin1993,bryden1999,lubow1999,jangcondell2012}. 
The gap radii (half the gap-width) for the planet models at $d_p=5$~AU, 10~AU, 20~AU, and 30~AU are 0.4~AU, 0.9~AU, 1.7~AU, and 2.6~AU, respectively.   The dust density within the gap has a floor-value at the gap midplane of $\rho_{\rm dust}=10^{-21}$~g~cm$^{-3}$ and is optically thin to the planet's infrared radiation, which is described in Section~\ref{sec:thermal}. The density within the gap is a factor of six orders of magnitude lower than the disk; however, we explore higher degrees of gap filling by dust in Section~\ref{sec:considerations}.
We emphasize that these models are designed to understand and isolate the local effects of planetary heating, and that we do not include variations in the 3D density structure near the planet, either excess flaring near the edges of the gap \citep[e.g.,][]{jangcondell2009}, or accretion streams, both of which will alter the vertical and radial disk structure and should be examined in future work (see also discussion in Section~\ref{sec:considerations}).  The density structure of the bulk disk is shown in Figure~\ref{fig:fullmod}a.  

\subsubsection{High-energy Stellar Irradiation Field}

We calculate the UV and X-ray radiation field from the central star for each of the four planet locations, $d_p$, from a single 2D slice and assume the high energy stellar radiation field is otherwise azimuthally symmetric.  The procedure for calculating the 2D X-ray and FUV field are the same as described in \citet{cleeves2013a,cleeves2014par}, using the Monte Carlo radiative transfer code of \citet{bethell2011code}. The FUV radiation is used to estimate the gas temperatures (S. Bruderer, in private communication), which are coupled to the dust temperatures in the region where the planet's impact is the greatest, and as such, our results do not depend on the gas temperature in the upper layers where the gas and dust temperatures are decoupled.  The disk is quite opaque to the FUV stellar photons (wavelengths between $930-2000$~\AA, including that of Lyman-$\alpha$) due to our use of an unsettled disk model.  A substantial amount of FUV continuum and line (Lyman-$\alpha$) radiation is down-scattered into the gap, aided by the UV-exposed, optically thick outer gap edge, which tends to deflect photons both into the gap and away from the disk.  However, the scattered continuum and line photons within the gap do not penetrate far into the gap edges and do not have a substantial impact on the chemistry beyond the illuminated wall.  The integrated FUV field between $\lambda=930-2000$~\AA\  is shown in Figure~\ref{fig:fullmod}b. 

The X-ray radiation field is computed using the same transfer code as the UV with the combined gas and dust X-ray absorption cross sections of \citet{bethell2011xr} and Thompson scattering.  Because the X-ray photons are stopped at larger column densities much further into the disk, the presence of the gap does not have as significant of an effect as it does for the FUV (Figure~\ref{fig:fullmod}c).  The gap exposes more of the disk to X-ray photons, but the overall X-ray field is much more sensitive to distribution of disk mass rather than the specific gap location; these Hill radius-sized gaps are relatively narrow compared to the path length of an X-ray photon at $\sim5$~keV.   For cosmic ray ionization, we use the model of \citet{cleeves2013a} for the cosmic ray ionization rate at solar maximum for the present day Sun ($\zeta_{\rm CR}\sim2\times10^{-19}$~s$^{-1}$ per H$_2$) as an upper limit to the contribution from cosmic rays, consistent with the upper limit determined from observations of the TW Hya disk \citep{cleeves2015tw}.  

\subsection{3D Thermal Structure}\label{sec:thermal}
We calculate the thermal structure of the disk assuming two passive radiation sources: the self-luminous central star the protoplanet shining due to its accretion luminosity.  We assume a circumplanetary accretion rate of $\dot{M}=10^{-8}$~M$_\odot$~year$^{-1}$ on an $M_p=1$~M$_{\rm Jup}$ and $R_p=3$~R$_{\rm Jup}$ planet, corresponding to an accretion luminosity of $L_{\rm acc}=5\times10^{-4}$~L$_\odot$.  In this simple picture, the planet is treated as a spherical, 1500~K blackbody that heats its self-carved gap from the inside.   The dust temperatures are calculated using the code TORUS \citep{harries2000,harries2004,kurosawa2004,pinte2009} assuming radiative equilibrium with the Lucy method \citep{lucy}.  We note that we do not include the differential rotation of the circumstellar disk in the dust temperature calculation because radiative equilibrium at $\sim50$~K is quickly attained within minutes to hours \citep{woitke1999}.  For a planet at 10~AU, this timespan is negligible compared to the $\sim30$~year period of the planet's orbit, thus planetary heating of the disk is expected to remain a local phenomena and will not become highly sheared out by differential rotation.
 The thermal structure for the bulk circumstellar disk (away from the planet) is shown in Figure~\ref{fig:fullmod}d for the planet at $d_p=20$~AU.  The large-scale physical properties are largely similar for all four planet locations in our models (i.e., the presence of the gap does not significantly change the thermal/irradiation structure away from the gap).   All four models have an outer disk radius of 50~AU and an inner disk radius of 0.2~AU. 
 
  \begin{figure*}[tbh!]
\begin{centering}
\includegraphics[width=0.74\textwidth]{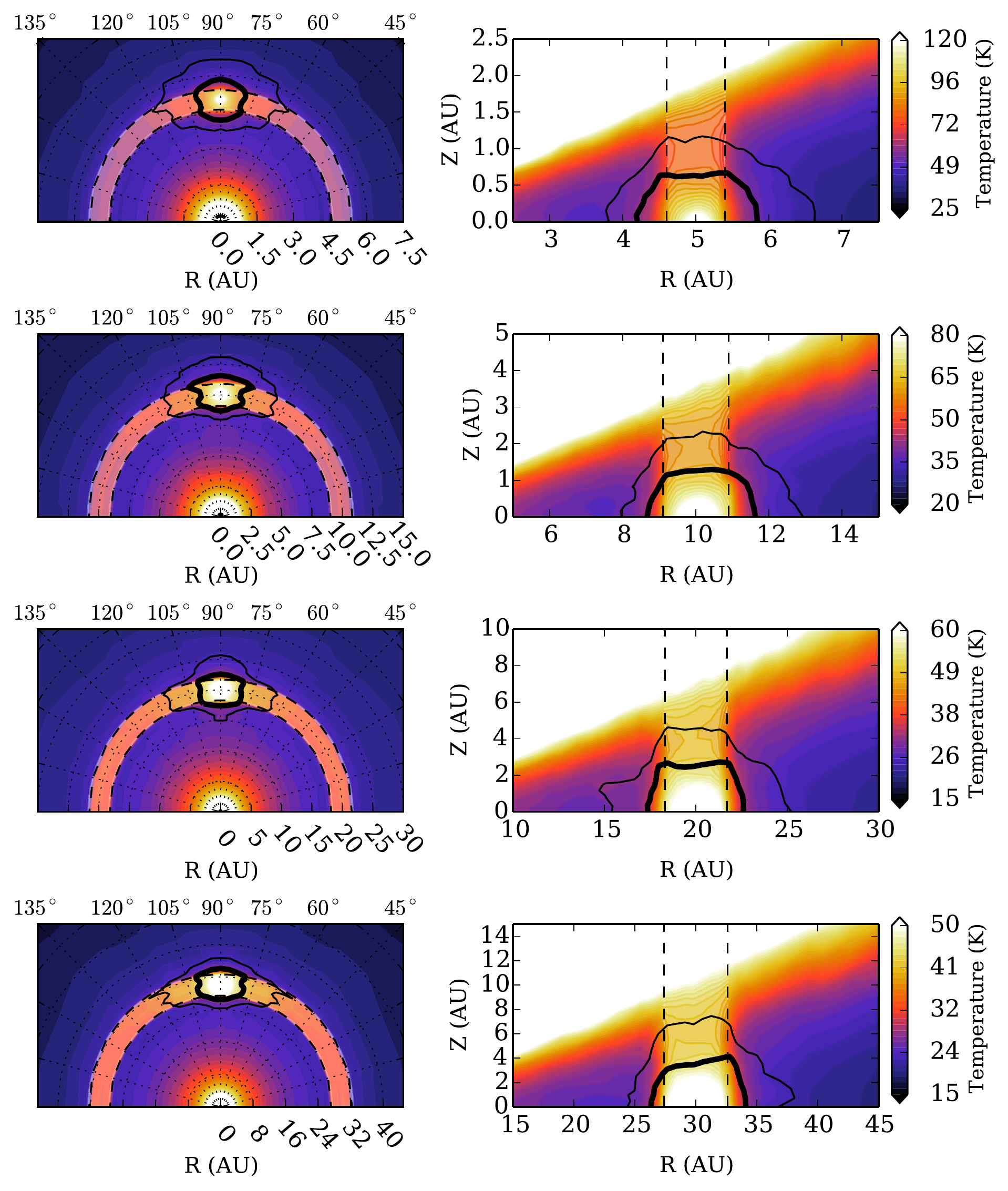}
\caption{Thermal structure in the region near the planet.  Left column shows the midplane temperature over radius and azimuth (planet is located at 90$^\circ$ in all cases).  Right column shows the vertical temperature structure centered on the planet. The thin (thick) contour line highlights the region of the disk where the planet increases the local temperature by $>10\%$ ($>30\%$). Top, middle and bottom rows are models for planets at 5~AU, 10~AU, 20~AU and 30~AU, respectively.  \label{fig:thermal08}}
\end{centering}
\end{figure*}
 
 The additional heating by the planetary companion primarily results in an azimuthally extended ($\Delta\phi\sim\pm30^\circ$) but somewhat thin ($R\lesssim5$~AU) swath of material along the gap edges, centered on the planet itself.  The temperature structures in the midplane and for a vertical slice centered on the planet's location are shown in Figure~\ref{fig:thermal08}.  
 The contours highlight the change in temperature due to the presence of the planet.  The  planet increases the temperature by greater than 10\% within about $\Delta R \sim 1.1$~AU, 1.6~AU, 2.7~AU, and 4~AU from the edge of the gap for the 5, 10, 20, and 30~AU orbital radius, respectively.  The absolute change in temperature at the gap edges closest to the planet ($\lesssim0.5$~AU from the wall) due to the additional heating corresponds to an increase from 42~K to 60~K ($d_p=5$~AU), 35~K to 47~K ($d_p=10$~AU), 27~K to 34~K ($d_p=20$~AU), and 24~K to 31~K ($d_p=30$~AU) for both gas and dust temperatures.  These substantial $\sim10-20$~K changes in cold, dense molecular gas will substantially alter the local chemistry close to the planet.

\subsection{Chemical Model}\label{sec:chemtech}
From the physical structure outlined above, we can estimate molecular abundances as a function of 3D position throughout the disk.  The disk chemical code used for the calculations is presented in \citet{fogel2011} and further expanded in \citet{cleeves2014par}.  The chemical code itself is inherently 1+1D, and so to address this limitation we extract 2D azimuthal cuts from full 3D model to compute the abundances, and reconstruct the full disk profile from the individual 2D calculations.  We consider ten 2D slices at $3^{\circ}$ intervals in azimuth ranging from the planet's location to $30^{\circ}$ away from the planet, and assume that the planet is symmetric upon reflection.  For the rest of the disk beyond $30^{\circ}$, we calculate the chemistry based upon one 2D slice on the opposite side of the disk from the planet, i.e. the ``anti-planet'' side.   Differential rotation is not included in the present models but will be explored in future work (see further discussion below).   The 1+1D \citet{fogel2011} code calculates the time-dependent abundances relative to hydrogen as a function of radius and height from the midplane based on a fixed grid of temperature, density, and radiation field conditions.  The chemical reaction network is based on the OSU gas-phase chemical network \citep{smith2004}, expanded to include grain-surface chemistry in the method of \citet{hhl1992}, where the grain-surface reactants ``sweep-out'' the surface at a rate related to the binding energy and mass of the reactant.   We treat the hydrogen binding energy in the method of \citet{cleeves2014wat}, where we assume that the binding energy for desorption processes is the chemisorbed value such that H$_2$ can form even at high temperatures, while the binding energy adopted for the rate of hydrogenation reactions  on the surfaces of cold dust grains is assumed to be the lower, physisorbed value, or $E_b({\rm H})=450$~K, such that the H-atoms are highly mobile across the ice mantle.   The reaction network has a total of 6292 reactions, including both chemical reactions and physical processes (i.e., ionization, desorption, etc.) and 697 species.  All chemical calculations are examined after 1~Myr of chemical evolution.

\section{Results}\label{sec:results}
In the following section, we present chemical abundance signatures due to the additional heating from a planetary companion as calculated over the disk.  We then simulate the emergent line emission for the fully 3D physical and chemical model.  Based upon these emission models we generate ALMA simulations of planets embedded in disks.  

\subsection{Chemical Abundance Results}\label{sec:chem}
The most important chemical effect from the planet is the thermal desorption of molecular species that are otherwise frozen out as ices in the midplane at the orbital distance of the planet, which we term ``primary tracers.''  As discussed in Section~\ref{sec:thermal}, the dust temperature rapidly equilibrates to the local irradiation conditions.  To assess whether the simplification of neglecting differential rotation is important for the chemical structure, we must compare the chemical timescales for adsorption/sublimation to the orbital time.    If the chemical timescales are longer than the orbital time, the chemical effects of the planetary heating within the gap will become sheared out over azimuth.  In contrast, short chemical timescales relative to the planet's orbital time indicate the relevant chemistry is able to adjust quickly, and thus ``follow'' the planet.

\begin{figure*}[ht]
\begin{centering}
\includegraphics[width=0.85\textwidth]{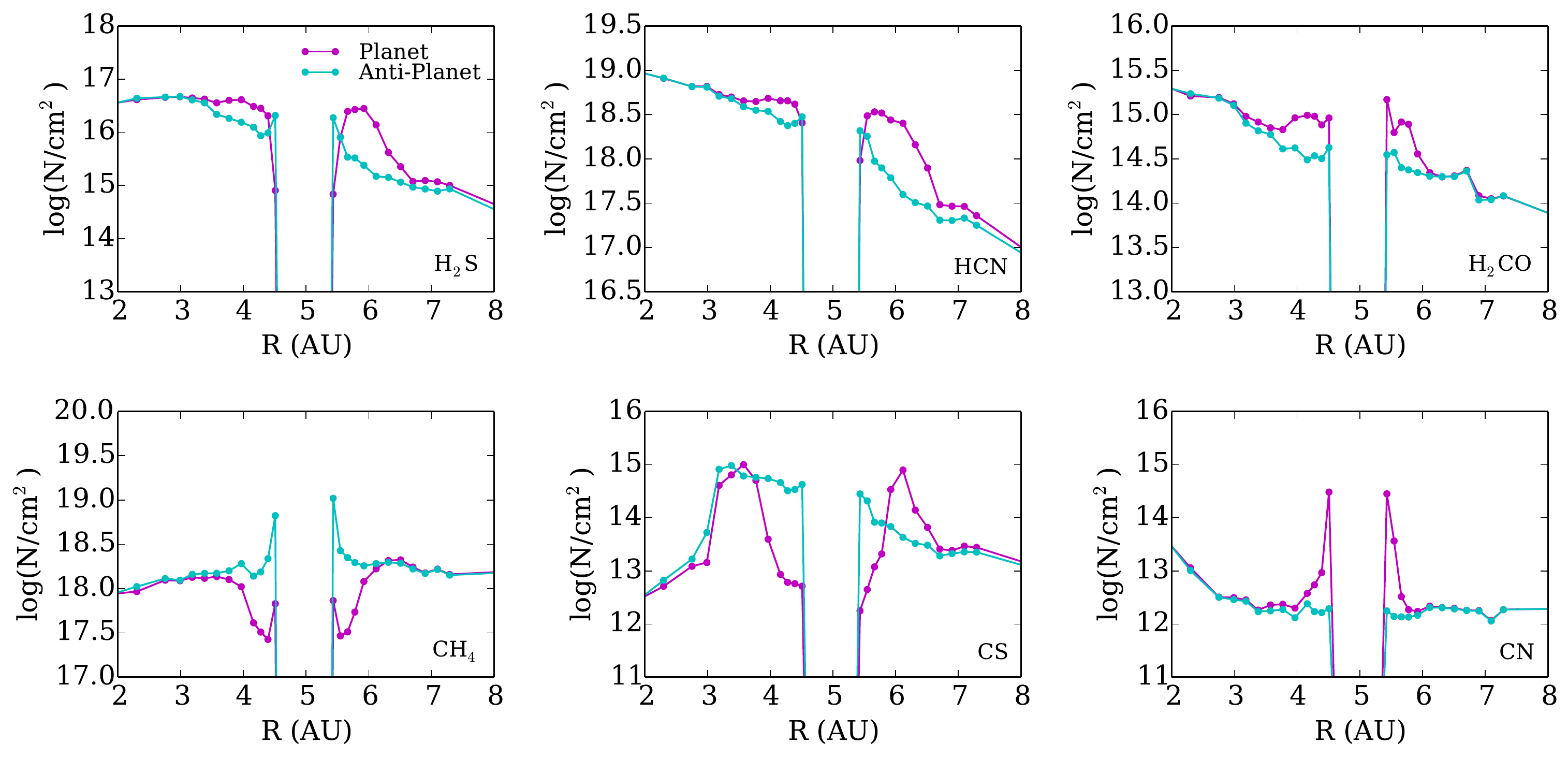}
\caption{Vertically integrated column densities of select molecular species (labeled in the lower right corner) after 1~Myr of chemical evolution for the planet at $d_p=5$~AU with an accretion luminosity of $L_{\rm acc}=5\times10^{-4}$~L$_\odot$.  The column density at the disk azimuthal angle centered on the planet (90$^\circ$ in Figure~\ref{fig:thermal08}) is shown in magenta, where the points correspond to the specific radii calculated in the chemical models.  The chemistry of the disk near the gap in the absence of the planet is shown in cyan, i.e., at the `anti-planet' side of the disk.  \label{fig:column_0508}}
\end{centering}
\end{figure*}

\begin{figure*}
\begin{centering}
\includegraphics[width=0.85\textwidth]{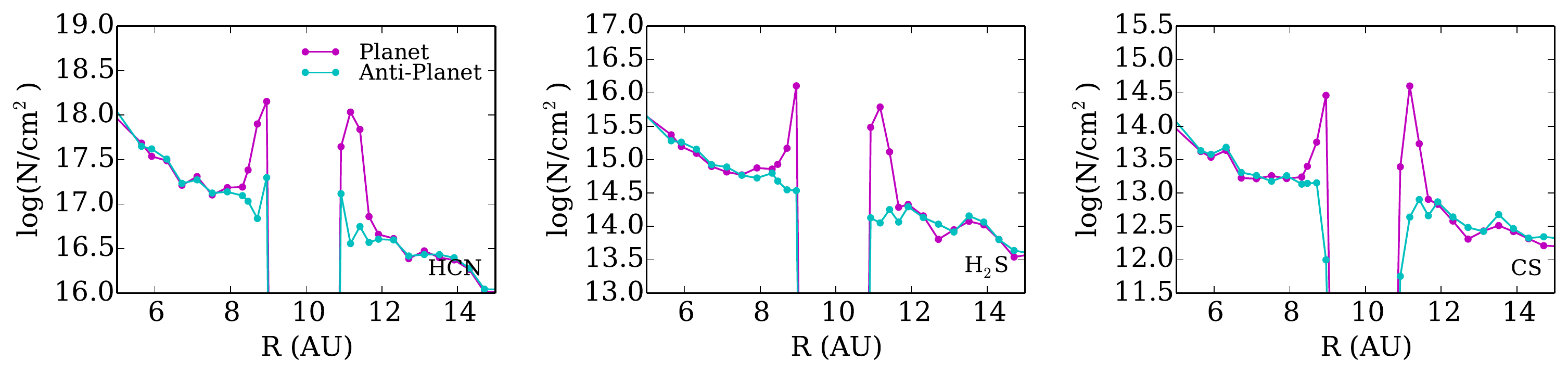}
\caption{Vertically integrated column density of select species for the planet at $d_p=10$~AU.  Figure as described in Figure~\ref{fig:column_0508}.  \label{fig:column_1008}}
\end{centering}
\end{figure*}

\begin{figure*}
\begin{centering}
\includegraphics[width=0.85\textwidth]{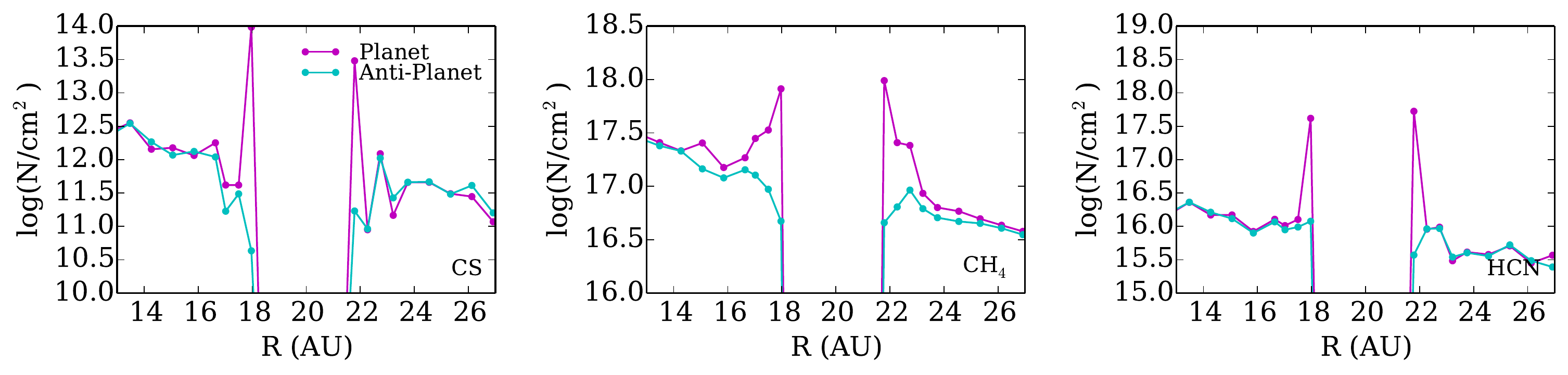}
\caption{Vertically integrated column density of select species for the planet at $d_p=20$~AU. Figure as described in Figure~\ref{fig:column_0508}. \label{fig:column_2008}}
\end{centering}
\end{figure*}

\begin{figure*}
\begin{centering}
\includegraphics[width=0.85\textwidth]{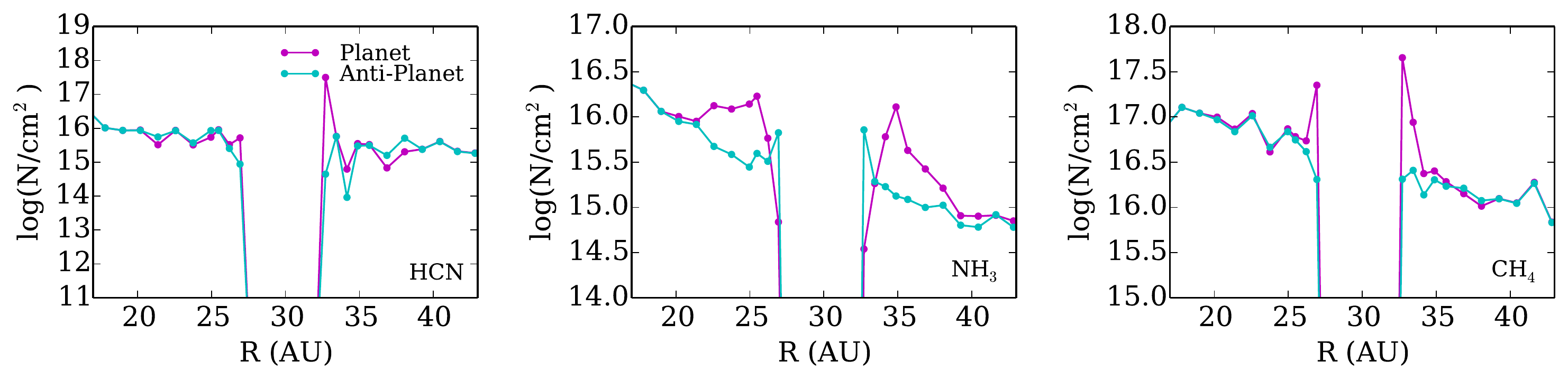}
\caption{Vertically integrated column density of select species for the planet at $d_p=30$~AU. Figure as described in Figure~\ref{fig:column_0508}. \label{fig:column_3008}}
\end{centering}
\end{figure*}

 In our model, the midplane at $d=10$~AU has a density of approximately $n_{\rm H_2}\sim10^{10}$~cm$^{-3}$ and a temperature of $\sim50-60$~K near the planet.  The timescale for freeze-out of a molecule is related to the surface area of grains per unit volume, or $n(\sigma_{\rm gr})$ cm$^2$ cm$^{-3}$, as well as the thermal speed of the molecule in the gas phase, $v_{\rm X}$, such that molecule  ${\rm X}$ freezes out in a characteristic time $t_{\rm fo} ({\rm X}) =\left (n(\sigma_{\rm gr}) v_{\rm X}\right )^{-1}$. At gas densities of $n_{\rm H}=10^{10}$~cm$^{-3}$, a typical grain surface area density is $n(\sigma_{\rm gr})\sim10^{-11}$ cm$^2$ cm$^{-3}$ (i.e., 0.1~$\mu$m-sized grains at an abundance relative to H-atoms of $6\times10^{-12}$), an H$_2$CO molecule (for example) with mass of 30~amu, will collide with a grain on average every $\sim0.3$~years, which is roughly the time for circumstellar disk chemistry to reset when not directly heated by the planet.  The corresponding timescale for thermal evaporation of an H$_2$CO molecule \citep[assuming a desorption temperature of $T_d\sim2050$~K;][]{garrod2006} on a 50~K dust grain based on the Polyani-Wigner relation \citep[see][]{fogel2011} is 0.02 years (the timescale for molecules to evaporate when exposed to the planet's heating).  Both of these timescales are sufficiently rapid compared to the orbital time around a 1~M$_\odot$ star for the planets considered here (about 11, 32, and 89 years) and therefore we expect the chemistry of the primary tracers to follow the planet and not experience strong azimuthal shear.  It is important to note that the relevant timescales will increase with height; however, in all cases, the chemical effect of the planet is limited to a narrow vertical region close to the midplane where $z/r<0.1$, over which the density only decreases by $\sim30\%$. 
 
 Molecules that form via gas-phase or grain-surface chemistry as a direct result of primary species desorption are ``second-order'' chemical effects -- i.e., secondary tracers.  The timescales for secondary species formed from primaries to return to the low-temperature state will depend on the particular formation pathways for the secondary species and the availability of He$^+$ ions to break up newly formed molecules that would otherwise not be present without the planet \citep[e.g.,][]{bergin2014,furuya2014}. In this case, the distribution of secondary species formed due to the planet are more likely to be affected by shear (see discussion below).

Given the size of the chemical network, we took an unbiased approach to search for promising chemical tracers of planetary heating, both primary and secondary.  We calculate the vertical column density of every species in the network versus radius, and filter out species with low column density near the planet, $N_X \le 10^{10}$~cm$^{-2}$, since these will be the most difficult to detect.  We then calculate the fractional difference between the column densities, $\Delta N_X(R)$,  at the azimuthal location of the planet and the anti-planet side, and integrate this quantity over radius to get an estimate for the total fractional change due to the planet.  We then sort by the fractional change to identify which species are most affected by the additional heating.  The results of this ``blind search'' for gas-phase tracers are shown in Figures~\ref{fig:column_0508} ($d_p=5$~AU), \ref{fig:column_1008} ($d_p=10$~AU), \ref{fig:column_2008} ($d_p=20$~AU), and  \ref{fig:column_3008} ($d_p=30$~AU).

For certain species, the additional heating changes the vertical column density at the edges of the gap by many orders of magnitude.  For most ``signpost'' molecules, the heating from the planet lifts the local dust temperature at the midplane from below the particular species desorption temperature to above, causing a large enhancement in the column density at the edges of the gap.  Two species stand out from this trend for the planet at 5~AU, where CS and CH$_4$ instead exhibit large deficits.  The assumed binding energies of CS and CH$_4$ are $E_b=2000$~K and $E_b=1330$~K, respectively.  In the case of CS, the 5~AU planet causes an initial, fast desorption of CS (and higher column density), but the heating simultaneously increases the abundance of gas-phase oxygen bearing molecules.  Over $\sim2000$~year timescales, the chemical network converts the CS to OCS ice, resulting in the net chemical deficit plotted in Figure~\ref{fig:column_0508}.  The orbital time at 5~AU is clearly much shorter than the chemical timescale for this process to occur, and so in this case, the conversion from CS to OCS (and the corresponding CS-deficit) is over-predicted in our simplified models.  To properly model the planet's effects on the CS chemistry in the inner disk requires models that include differential velocity shear to address these second-order time-dependent effects.  However, for planets located further out in the disk, CS shows simple net desorption behavior without further chemical processing for the more distant, 10 and 20~AU planets.  The change in chemical behavior arises because the chemistry leading to OCS is not as efficient radially further out where the gas-phase oxygen is depleted.  As a result, CS becomes a ``primary'' planet tracer for the planets located further out.  Thus using thermal desorption as a search tool for planets is more reliable in the more chemically inactive outer disk.   

For CH$_4$, this species' low binding energy ($T_d\sim27$~K) allows for it to be in the gas phase regardless of the planet's location.  The deficit in CH$_4$ is purely a second-order effect, where the CH$_4$ at 5~AU primarily forms via gas-phase channels originating from CH$_5^+$.  The presence of additional, sublimated gas-phase species that are also able to  react with CH$_5^+$ will form other molecules besides CH$_4$.  These additional pathways stymy the formation efficiency of CH$_4$ and leads to the CH$_4$ deficit in Fig.~\ref{fig:column_0508} of about a factor of three in column density.  This process likewise occurs over much longer timescales than the planets' orbital times and should be studied in further detail including disk differential rotation to fully characterize the CH$_4$ chemistry in the presence of the planet.

There is some variation of the radial extent of the region affected by the planet as seen in the column density plots.  For example, as seen in Figures~\ref{fig:column_2008} and \ref{fig:column_3008}, the 20~AU and 30~AU planets exhibit  enhanced HCN confined to a narrow region close the gap edges ($\lesssim0.6$~AU), while CH$_4$ is enhanced over a much wider region, $\lesssim1.5$~AU.  The main factors that determine the chemically affected region are the gap size \citep[i.e., a larger gap reduces the amount of heating at the wall's interface;][]{cleeves2011} and the binding energy of the species in question.  CH$_4$ is more weakly bound to the grain surface, $E_B\sim1360$~K, compared to HCN, $E_B\sim1760$~K \citep{hasegawa1993}, and so the region near the planet is only sufficiently hot to evaporate HCN close to the gap walls.  

In addition to HCN and CH$_4$, NH$_3$ is also enhanced in the presence of the 30~AU planet (see Fig.~\ref{fig:column_3008}).  This behavior is a feature of the relatively low NH$_3$ binding energy used in the present models, $E_b=1100$~K \citep{hasegawa1993}, corresponding to a desorption temperature of $T_d\sim28$~K.  Alternatively, NH$_3$ deposited on a H$_2$O ice surface has a much higher binding energy, $E_b~3200$~K \citep{collings2004}, or $T_d\sim90$~K.  Thus if the higher desorption temperature applies, NH$_3$ evaporation may be a more useful tracer of planets in the inner few AU of a cool protoplanetary disk, or further out for warmer disks like those around Herbig Ae/Be stars.  In summary, the ideal chemical tracer for identifying planets is fundamentally a balance of the appropriate chemical binding energies and the disk thermal structure.  

\begin{figure*}
\begin{centering}
\includegraphics[width=0.74\textwidth]{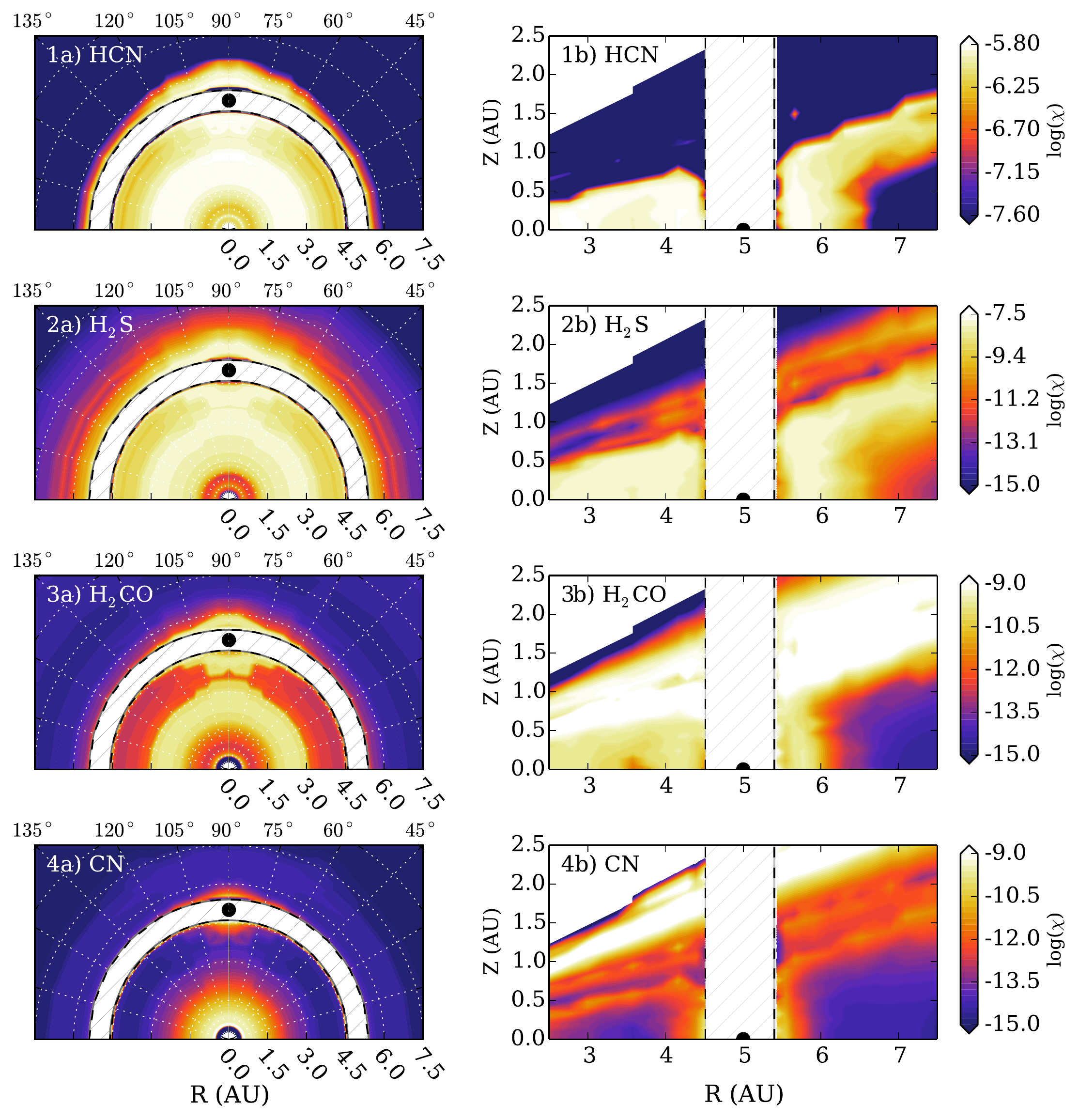}
\caption{Chemical abundances for species that are sensitive to the planet's additional heating.  The left column shows the abundances in the disk midplane, while the right column shows the vertical abundances centered on the planet.  Chemical models are shown for the planet at $d_p=5$~AU. \label{fig:tallmidplane_0508}}
\end{centering}
\end{figure*}

\begin{figure*}
\begin{centering}
\includegraphics[width=0.74\textwidth]{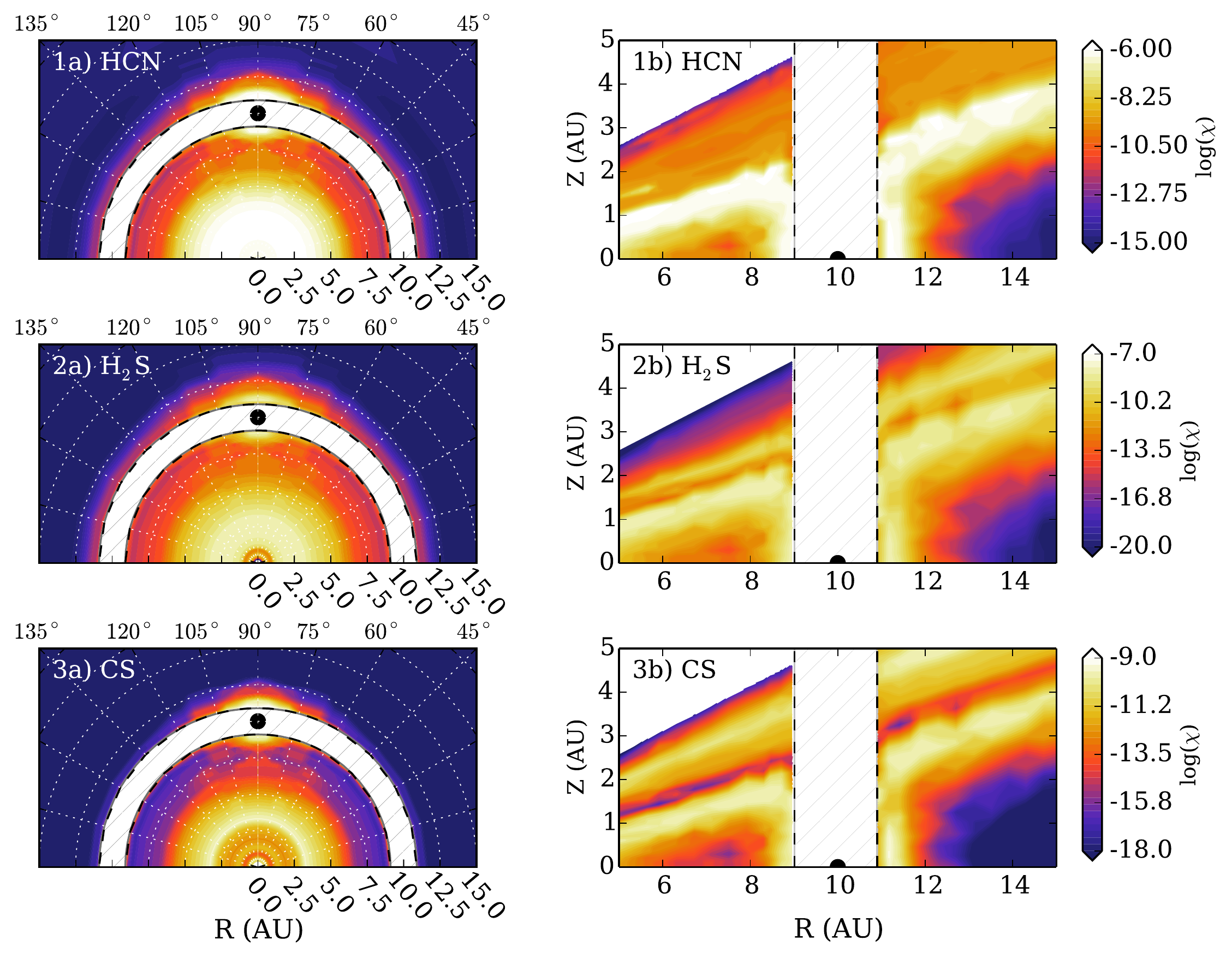}
\caption{Chemical models for the planet at $d_p=10$~AU.  Panels are the same as for Figure~\ref{fig:tallmidplane_0508}.\label{fig:tallmidplane_1008}}
\end{centering}
\end{figure*}

\begin{figure*}
\begin{centering}
\includegraphics[width=0.74\textwidth]{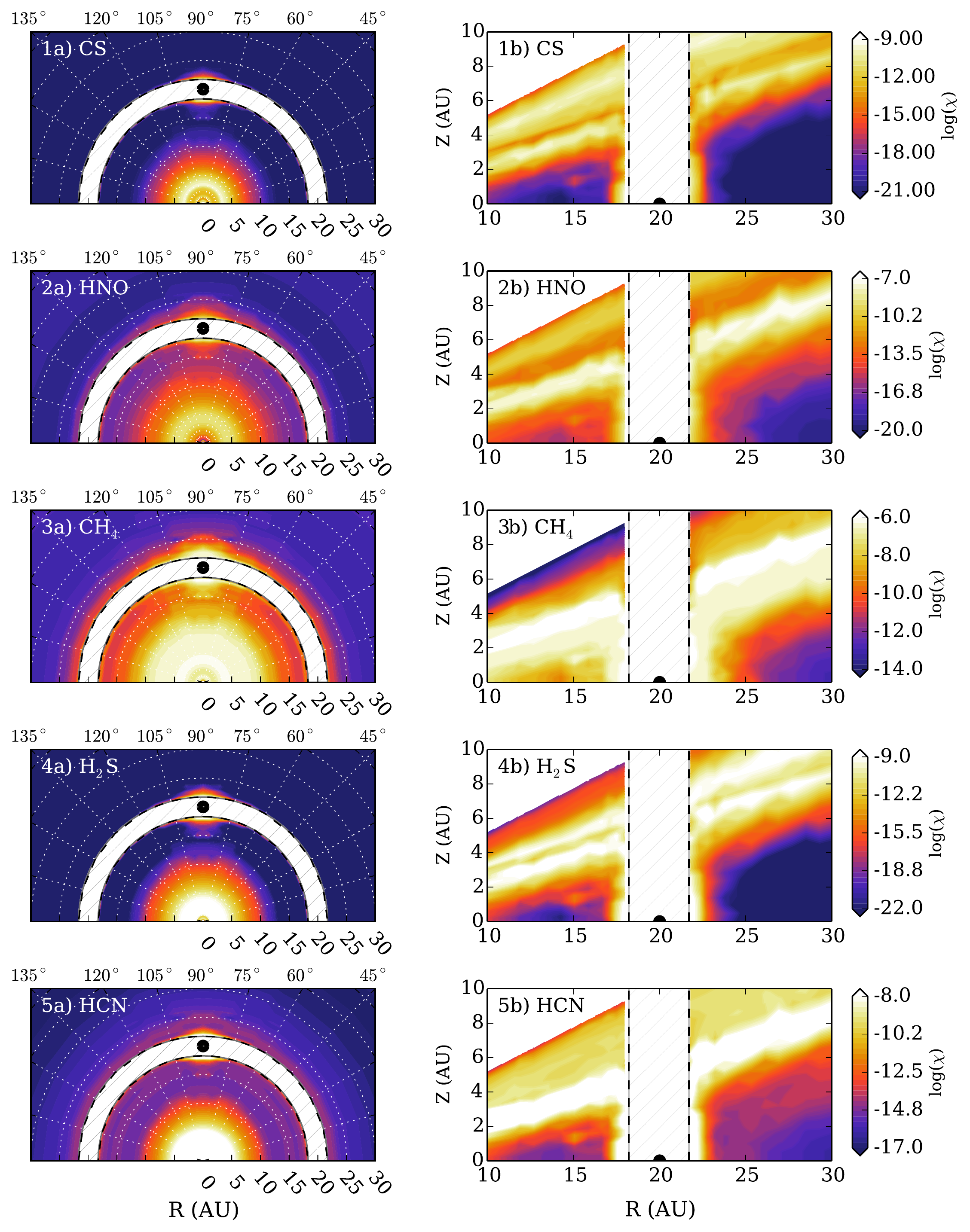}
\caption{Chemical models for the planet at $d_p=20$~AU. Panels are the same as for Figure~\ref{fig:tallmidplane_0508}.\label{fig:tallmidplane_2008}}
\end{centering}
\end{figure*}

\begin{figure*}
\begin{centering}
\includegraphics[width=0.74\textwidth]{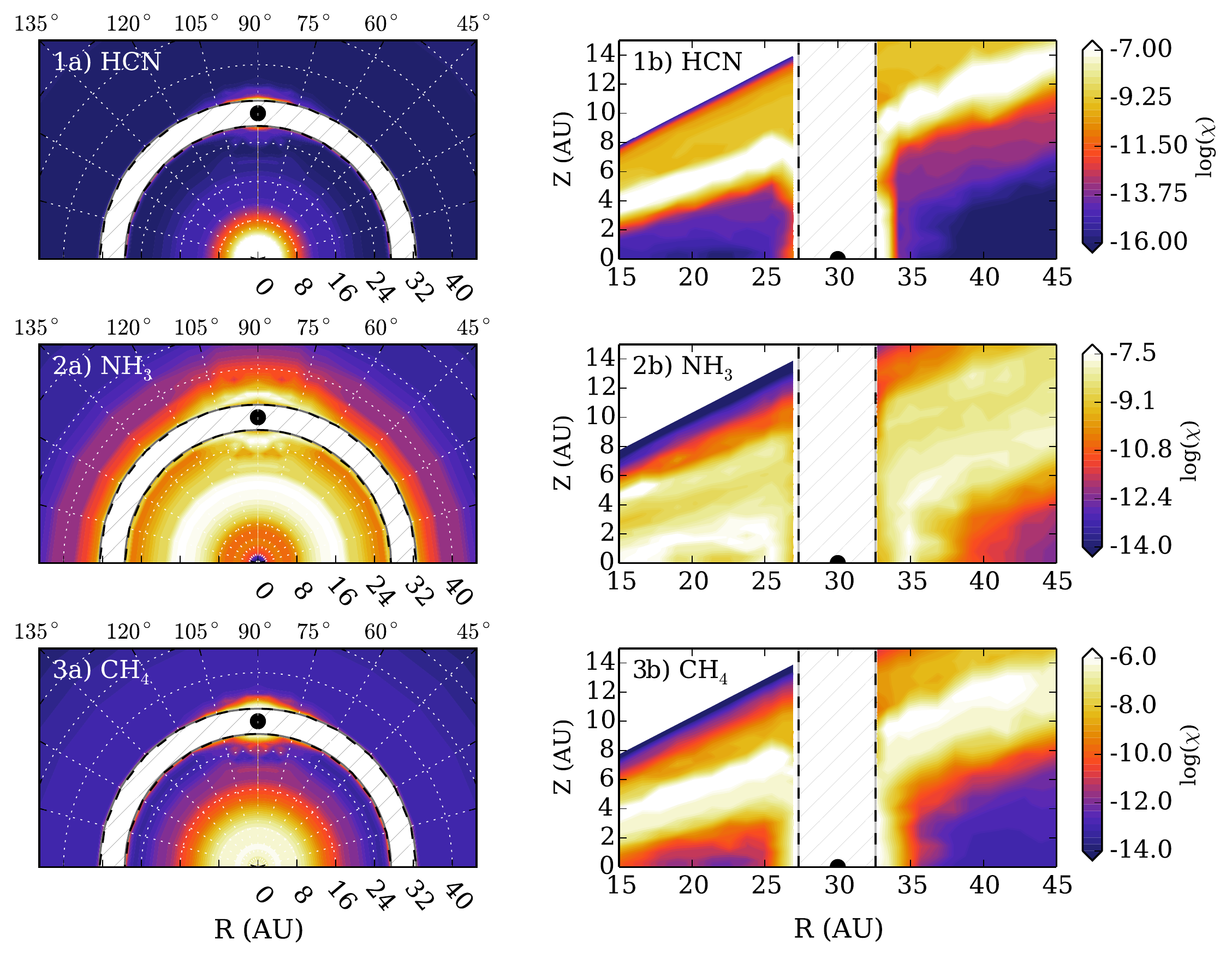}
\caption{Chemical models for the planet at $d_p=30$~AU. Panels are the same as for Figure~\ref{fig:tallmidplane_0508}.\label{fig:tallmidplane_3008}}
\end{centering}
\end{figure*}

From the 2D chemical model slices we reconstruct the 3D abundance profiles.  We plot the abundance structure for particularly promising species for the four planet locations in Figures~\ref{fig:tallmidplane_0508}, \ref{fig:tallmidplane_1008}, \ref{fig:tallmidplane_2008}, and \ref{fig:tallmidplane_3008}.  From these plots, it is clear that the chemical effects extend over a larger azimuthal range than radial range for most cases.  The abundance enhancement furthermore spans the full wall height until the abundance profile merges into that set by the surface chemistry, as driven by the central star.  However, the vertical density structure is such that the densities are highest closest to the midplane and most tenuous near the surface.  Thus the midplane heating by the planet can have a large effect on the total, integrated column density, similar to the direct midplane illumination of transition disks by the central star \citep{cleeves2011}. 

Of all the species considered here, HCN is a particularly robust tracer with a simple midplane chemistry, such that the thermal effects from the planet are primarily its desorption.  The binding energy of HCN in our models is set to 1,760~K \citep{hasegawa1993}, i.e, a characteristic desorption temperature of $\sim44$~K .  For the planets beyond 10~AU, HCN should otherwise be frozen-out in the midplane for the majority of the disk.   Even at 5~AU (midplane temperature of $\sim40$~K), HCN is just beginning to freeze-out in the absence of the planet.  Thus HCN should be an excellent tracer of planets at orbital distance of $d_p>5$~AU from the central star.  Even if the particular HCN binding energy is revised, the location HCN as a planet-tracer will scale accordingly based upon the midplane temperature profile of the disk and the molecules' binding energy to the grain.  Furthermore, if the HCN binding energy varies with the properties of the substrate \citep[e.g.,][]{collings2004}, the azimuthal region away from the planet can be used as a baseline against which the HCN enhancement can be compared independent of the specific binding energy assumed.  For all four models, the enhancement in HCN is typically an order of magnitude in column density due to the planetary heating. In the absence of the planet's effects, the baseline HCN column density on the opposite side of the disk is $N_{\rm HCN} \sim 10^{16}$~cm$^{-2}$ for the planet at 30~AU and  $N_{\rm HCN} \sim 10^{18}$~cm$^{-2}$ for the planet at 5~AU.  The  column of HCN in the absence of the planet arises almost entirely from the HCN present at the warm disk surface.

\subsection{Line Emission}\label{sec:emission}
Many of these ``signpost'' molecular species have strong rotational transitions that can be used to observationally identify and characterize the local physical conditions near the planet.  Because of its simple interpretation and large column density, we focus on HCN emission as a tracer in the present paper, but emphasize that other, perhaps stronger tracers may exist, and the particular tracer will depend on the luminosity of the central star, which will set the thermal structure of the midplane at the disk radii probed by ALMA.  

We use the flexible line modeling code LIME \citep{brinch2010} to simulate the emergent line emission from the full disk, oriented face-on at a distance of $d=100$~pc.  We present HCN emission results for the planets at $d_p=10$~AU, 20~AU and 30~AU as the HCN signature for the 5~AU planet was not detectable or distinguishable from the background disk in any of the models considered. Because LIME samples the model's native grid by selecting random points and accepting/rejecting them based upon particular criteria (in our case, normalized density and HCN abundance), the code can potentially not adequately sample relatively small scale local features in the disk when considering the full 3D volume of the disk. Arbitrarily increasing the point number, however, will add more points into the inner, high density regions of the disk that are already well-sampled.  To address this issue, we cast two grids of points, a large grid over the full disk range (including the planet), and a refined spherical grid centered on the planet that contains $5\%$ of the total number points ($\sim$300,000) used to sample the model.  This insures that a sufficient number of points encompass both the disk and the region near the planet.  We furthermore tested this technique for a model without any planet (but still with the refined local grid) to confirm that the refined grid does not introduce emission signatures.  Such sub-grids will be useful in modeling all types of substructure with LIME. 

Because HCN is fairly optically thick at the surface, we model both the HCN and H$^{13}$CN isotopologue assuming an isotope ratio of $^{12}$C/$^{13}$C ~$=60$. We emphasize that at the high densities where the planet is depositing its heating that selective photodissociation effects on N$_2$ relating to HCN chemistry \citep{heays2014} are not expected to play a major role.   For the emission calculations, we consider the $J=4-3$ and $J=8-7$ rotational transitions for both species, which we identify as likely strong transitions accessible with ALMA using preliminary RADEX\footnote{http://home.strw.leidenuniv.nl/~moldata/radex.html} models \citep[RADEX is a statistical equilibrium solver under the assumption of the large velocity gradient approximation;][]{vandertak2007}. 
Using LIME, we calculate the line optical depth for the four transitions considered. For both HCN lines, the disk is optically thick ($\tau>100$) at all positions at the line center.  However, in the line wings at velocities $\delta v > 0.3$~km~s~$^{-1}$, the circumstellar disk is optically thin for both transitions, while the emission near the planet is thick.  Thus these lines will be useful for constraining the temperature of the emitting circumplanetary medium.  The H$^{13}$CN $J=4-3$ line is also thick at line center across the disk ($\tau~15-50$), but becomes thin and the lines, and has an optical depth of $\tau\sim1$ near the planet. H$^{13}$CN $J=8-7$ is marginally optically thick at line center ($\tau\sim10$), and thin at all locations in the wings.

\begin{figure*}
\begin{centering}
\includegraphics[width=1.0\textwidth]{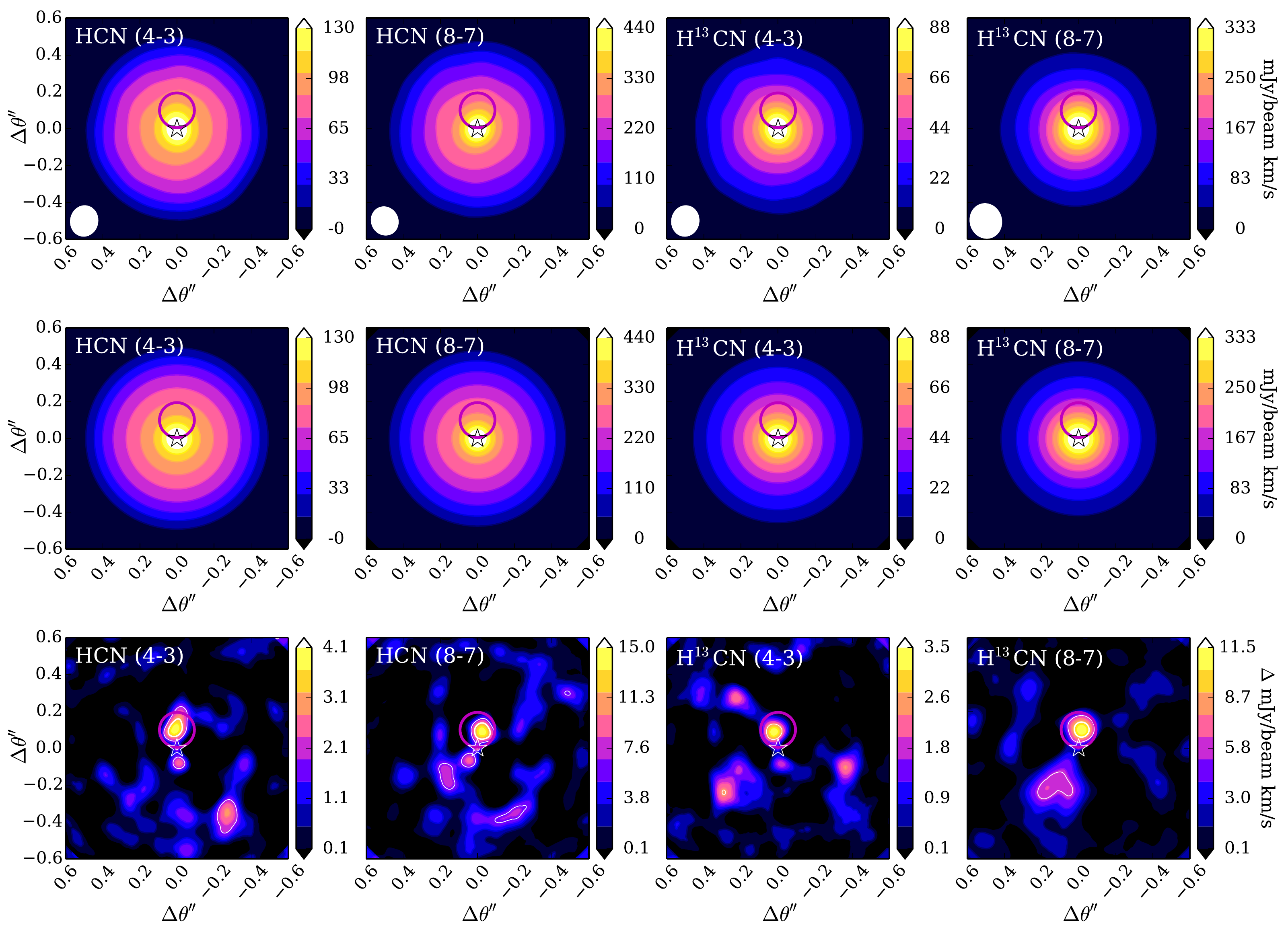}
\caption{Select emission lines for the planet at $d_p=10$~AU. The top row shows 10$h$ simulated observations of the indicated species as velocity integrated line emission. The middle row plots the annular averaged emission profile, that is representative of the background disk.  The bottom row shows the residual between the full disk simulation (top) and the averaged profile (middle), highlighting the asymmetric emission.  Contours for the residual plots are $3\sigma$ and $5\sigma$ (where applicable).  The magenta circle highlights the location of the planet. \label{fig:em_1008}}
\end{centering}
\end{figure*}

\begin{figure*}
\begin{centering}
\includegraphics[width=1.0\textwidth]{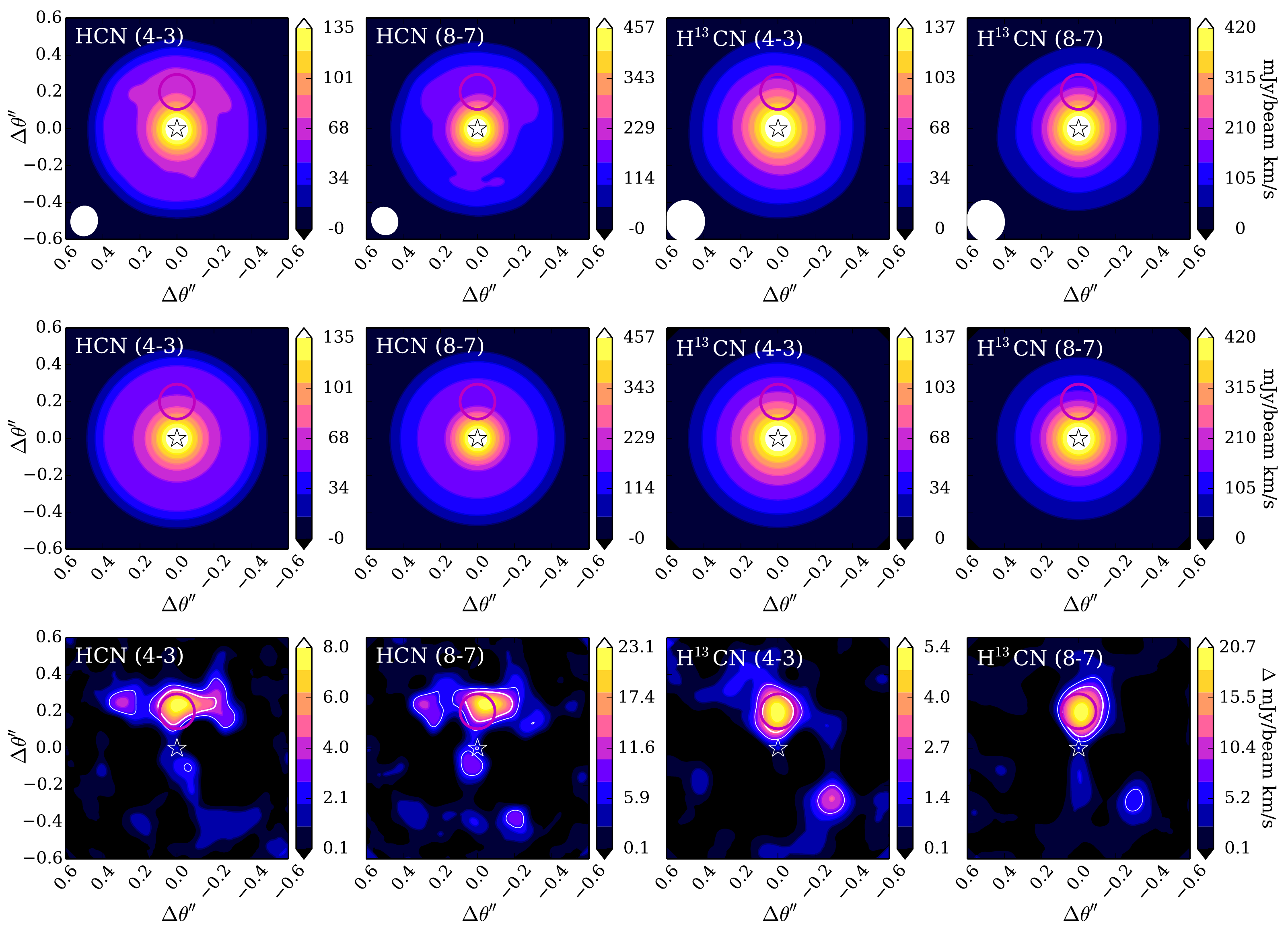}
\caption{Select emission lines for the planet at $d_p=20$~AU.  Panels are the same as for Figure~\ref{fig:em_1008}. \label{fig:em_2008}}
\end{centering}
\end{figure*}

\begin{figure*}
\begin{centering}
\includegraphics[width=1.0\textwidth]{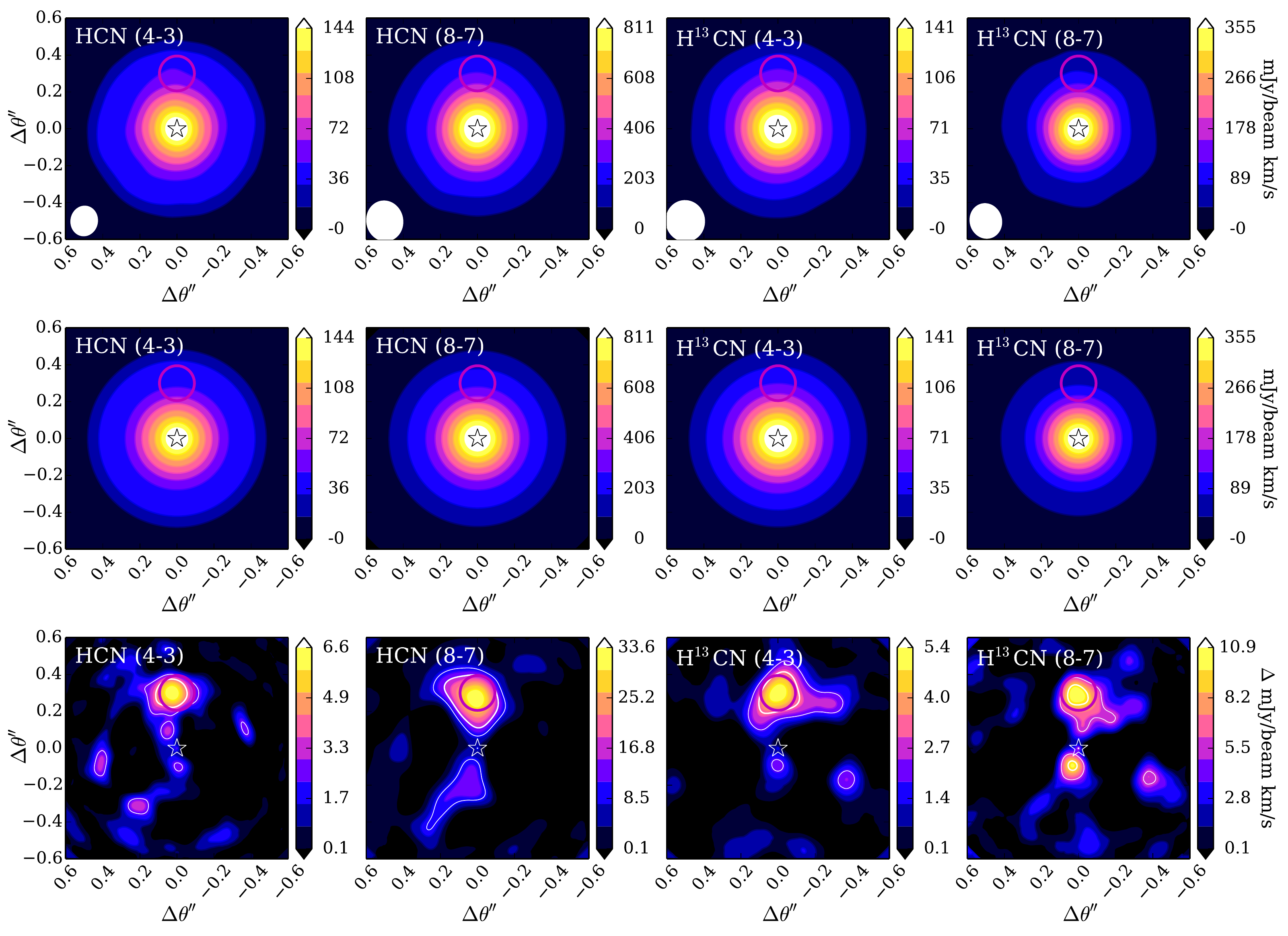}
\caption{Select emission lines for the planet at $d_p=30$~AU.  Panels are the same as for Figure~\ref{fig:em_1008}. \label{fig:em_3008}}
\end{centering}
\end{figure*}

   The LIME models represent ``perfect'' images of the emission lines without any thermal noise or beam-convolution (besides the inherent limitations of the pixel size, which is $0\farcs0025$ per pixel or 0.25 AU in our models).  To create more realistic emission models, we take the LIME output and, using the simulation capabilities of CASA\footnote{http://casa.nrao.edu/} we compute simulated ALMA observations in the same method of \citet{cleeves2014par}.  The alma.out13 antenna configuration is used for the $d_p=10$~AU simulation of both $J=4-3$ lines and the alma.out10 antenna configuration was used for the $J=7-6$. For the planet at $d_p=20$~AU (30~AU), alma.out16 (alma.out14) was used for $J=4-3$ and alma.out.11 (alma.out09) for $J=7-6$.  Configurations were chosen to provide optimal sensitivity and signal-to-noise to detect the planet.  We add realistic thermal noise with 0.25~mm precipitable water vapor, and then reconstruct the image using CLEAN applied with natural weighting.  The synthesized beam of the simulated observations is typically between $0\farcs1$ and $0\farcs2$.  The results of the simulated line observations (with noise) are shown in Figure~\ref{fig:em_1008} for the $d_p=10$~AU planet, Figure~\ref{fig:em_2008} for the $d_p=20$~AU planet, and Figure~\ref{fig:em_3008} for the $d_p=30$~AU planet.  
   
 In most cases, the emission from the planet is difficult to disentangle from the emission originating from the disk. However, the planet's sublimation effects are confined to a $\pm30^\circ$ azimuthal extent, which leaves the remaining  $\sim300^\circ$ as a counterpoint against which we can compare.  To better isolate the signature of the planet (without making use of a priori  knowledge of the planet's location), we average the emission from the entire disk in annuli to estimate the radial line emission profile of the circumstellar disk (Figure~\ref{fig:em_1008}, \ref{fig:em_2008}, and \ref{fig:em_3008}, middle row).   Then, using the average disk profile, we subtract off the smooth background component from the simulated observation and examine the residual features.   This technique will only work if the asymmetric feature is small and/or weak relative to the disk's emission, otherwise a median or averaging azimuthally restricted annuli (to mask out the feature) should be used instead.  Though this system is face-on, the technique would similarly work for systems that are not face-on if the inclination is known, i.e., by taking a projected annular average. 
 
 The bottom row of Figures~\ref{fig:em_1008}, \ref{fig:em_2008}, and  \ref{fig:em_3008} shows the residual emission, where the planets location is circled in all of the panels.  Without the additional knowledge of the planet's location, the signature in the $J=4-3$ is not strong enough to identify emission from the noise for the planet, primarily because of the high opacity of this line, which partially hides the deep emission from the midplane.  The lower opacity for $J=8-7$ for both isotopologues, however, makes for a substantially stronger signal from the planet, even though the noise is substantially higher at these frequencies.   The main limitation is the extremely high sensitivity required to measure the small fractional change in the emission signature from the bright circumstellar disk emission.

Because in our models the HCN abundance is highest along the walls of the gap near the planet, the face-on inclination provides the largest column density due to the planet.  However, if a more inclined viewing geometry may allow direct imaging of the gap wall depending on the overall scale height of the disk.  We examined emission simulations of a disk oriented at $60^{\circ}$ with the planet at 90$^{\circ}$ from the projected rotation axis (in the gap corner), and were unable to recover any emission from the planet.   If the planet is on the far-edge of the disk such that the outer gap wall is more directly viewable, and the emission signature may be observable but only for a fraction of the orbit.  Thus low inclination (face-on) disks are favorable for planet searches with chemistry.

\section{Further Considerations}\label{sec:considerations}
To simulate the 3D chemical models presented here, we have made a number of simplifying assumptions that we discuss further here. The assumed accretion rate for our models is on the high end of the range for quiescent planetary accretion as predicted by models, $\dot{M}=10^{-10}-10^{-8}$~M$_\odot$~year$^{-1}$, and is assumed to be constant. We found that it was difficult to detect the chemical signature of the planet for lower circumplanetary disk accretion rates, $\dot{M}\le10^{-9}$~M$_\odot$~year$^{-1}$. However, we emphasize that the corresponding accretion luminosity of the planets in our models, $L_{\rm acc}=5\times10^{-4}$~L$_\odot$ is indeed similar to that observed for the tentative LkCa~15 planet.  Furthermore, if young planets undergo frequent accretion outbursts \citep[increasing the planet's luminosity by many orders of magnitude, see][]{lubow2012} and if the cadence of such outbursts is competitive with the chemical timescales, the region affected by the planet may be much larger and more readily detectable. 

One significant simplification in the present models is the use of a constant gas-to-dust mass ratio of 100. Disks are observed to have a deficit of small dust grains in their upper layers, attributed to the settling of grains from the surface into the midplane \citep{furlan2006}.  One of the overall effects of settling is to create a warmer disk, where stellar radiation penetrates deeper.  If the disk becomes too warm, ice sublimation due to the planet may be less effective and one would have to i) observe gas-phase tracers with relatively higher binding energies to achieve the same contrast between the planet and the background disk or ii) look for planets at larger radii where the circumstellar disk is cooler. The addition of dust mass (in particular dust grain surface area) in the midplane will also increase the rate of absorption onto grains (i.e., decrease the freeze-out time) and as a consequence, this may help reduce the effects of shear by disk rotation.  On the other hand, if small grains are quickly converted to large grains with low surface area-to-mass ratios, then the freeze-out time will increase, and there may be a lag in the removal of planet-desorbed molecules from the gas phase thus spreading out the thermal effect of the planet over a larger azimuthal area (see Section~\ref{sec:conclusions}). 

\begin{figure}
\begin{centering}
\includegraphics[width=0.485\textwidth]{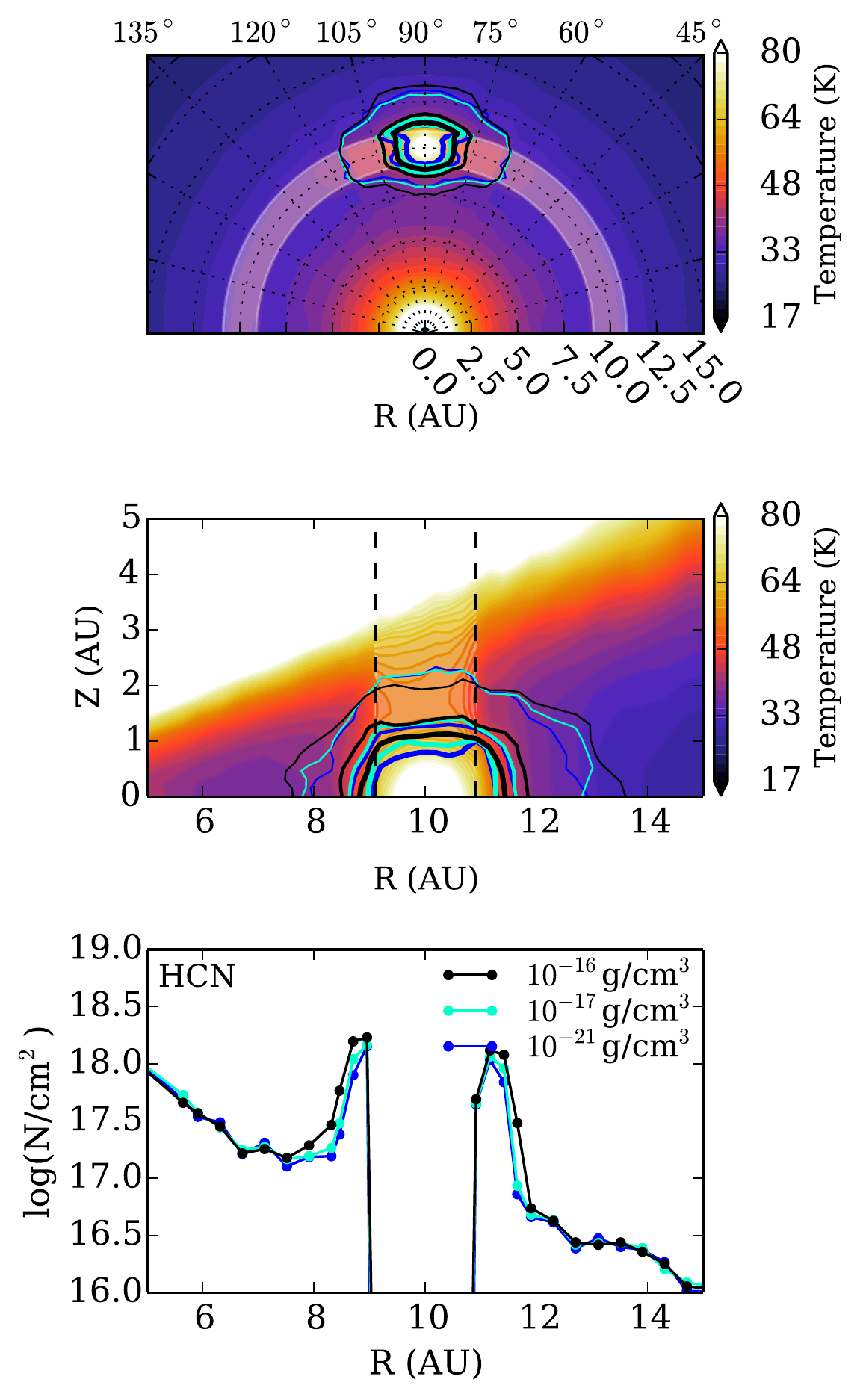}
\caption{  Partially filled gap models. The top and middle panels show the azimuthal and vertical temperature structure, respectively.  The bottom panel shows the HCN column density at the azimuthal location of the planet, where maximum heating is deposited. The fiducial model is shown as the blue contours in all panels ($\rho_{\rm dust}=10^{-21}$~g~cm$^{-3}$). Models where we have increased the dust density in the gap by four orders of magnitude (cyan) and five orders of magnitude (black) are indicated. \label{fig:fill}}
\end{centering}
\end{figure}

We have also used a simplified density model in the present calculations. Recent work by \citet{dong2014} has shown that the dust density distribution within the gap is different for small (sub-micron) and large (mm) sized grains.  Furthermore, the small grains readily fill the gap while the gaps are clear of the large grain population. In that work, the authors find a dust density within the gap of $\rho_{\rm dust}=10^{-17}$~g~cm$^{-3}$, which is four orders of magnitude higher than our simple, relatively empty gap model.  To explore how the addition of dust within the gap affects our results, we have calculated models for the planet at 10~AU where we have increased the minimum dust density within the gap by four and five orders of magnitude to simulate filled-gaps.  The enhanced dust models have gap densities that are similar to and above that predicted in \citet{dong2014}, and the dust density drop within the gap is a factor of 10 and 100 below a correspondingly gapless disk.  The results of this modeling is shown in Figure~\ref{fig:fill}. The most remarkable feature of these models is that the thermal structure of the protoplanetary disk near the young planet is similar regardless of the amount of dust within the gap. Moreover, the surrounding protoplanetary disk's temperature slightly increases with increasing dust in the gap.  The reason for this behavior is that some fraction of the thermal radiation emitted from the protoplanet is able to more freely escape in the vertical direction for the low density gap case and is accordingly lost to space.  Increasing the dust within the gap naturally traps more of the planet's heating, which is then reradiated isotropically, sending more heat into the protoplanetary disk that would otherwise be lost. We find a small increase in the HCN column corresponding to the slight increase in protoplanetary disk heating by this effect. The overall chemical structure is relatively unchanged.

 Another structural simplification made in this work was the assumption of axisymmetry. Massive planets will gravitationally interact with the circumstellar disk, generating density waves that can form spiral arms, changing the disk's radial and azimuthal density structure \citep[see the review of][]{kley2012}.  Planets can also excite vortices, creating large azimuthal pressure traps, observationally seen as asymmetries in the dust distribution \citep[as seen in IRS~48;][]{vandermarel2013}. Furthermore, the presence of the gap exposes the top of the outer gap edge to direct stellar irradiation, which enhances heating near the gap at the disk surface \citep{jangcondell2012}. While the latter effect is azimuthally symmetric, the heating can be locally exacerbated by the presence of a massive planet perturbing the vertical disk structure, creating local hot/cold spots near the stellar-irradiated surface at the planet's location \citep{jangcondell2009}.  We emphasize that all of these planet induced structures/physical effects can happen in concert with planet-heating and ice-desorption in the midplane.  The presence of spiral arms can be potentially confirmed as planetary in origin if associated with a warm accreting source. Furthermore, the stellar gap-edge heating occurs primarily at the surface of the disk, away from the planet's primary effects near the midplane, such that the use of spectral lines with high critical densities may help mitigate observational confusion.  Finally, the concentration of dust in pressure bumps can alter the chemical timescales such as the freeze-out time, though it should be noted that the small grains (which dominate the freeze-out given their high surface area to mass ratio) are still azimuthally distributed in IRS~48, providing a substrate for freeze-out.  Thus ideally one would find evidence of planets through multiple signatures, both chemical, physical, and thermal \citep[e.g., by detecting circumplanetary disk continuum emission;][]{zhu2015}.

Finally, we have assumed very simple adsorption/desorption physics, with a single HCN binding energy.  In reality, the HCN ice is expected to be mixed with less volatile ices, such as H$_2$O or CH$_3$OH.  Trapping can thus increase the desorption temperature to that of the strongly bound ice, $\sim5000$~K \citep[e.g.,][]{sandford1988,Sandford:1990fz,collings2003,collings2004}, potentially reducing the effects predicted in this work.  How trapping mediates desorption will essentially depend on how ices were initially deposited on the grain surface substrate -- simultaneously or sequentially, the thickness of the ice, whether the ice is amorphous or crystalline, which impacts the diffusivity of the trapped species \citep[e.g.,][]{collings2003}, and the specific molecule in consideration, and whether it is ``CO-like'' or ``H$_2$O-like.'' Ices like HCN are expected to be intermediate species, where some fraction directly sublimates and some fraction remains trapped, while, for example, CS, is expected to desorb more readily upon heating  \citep{viti2004}. While incorporating the detailed physics of the desorption process is beyond the scope of the paper, they should be taken into consideration when comparing to the inherently more complex observations. We note that if HCN is partially trapped (making for a weaker but existent signal), the technique of comparing the local (planet) emission signature against the disk-averaged emission will nonetheless help compensate for more complex chemical effects than considered in the present work.

\section{Conclusions}\label{sec:conclusions}

We have examined the chemical structure of simple toy models of a protoplanetary disk heated by two sources, the central star and an embedded, luminous young proto-Jupiter accreting at $\dot{M}=10^{-8}$~M$_\odot$~year$^{-1}$.  Chemical species with intermediate freeze-out temperatures, around $\sim40$~K, will be particularly useful tracers of the additional planetary heating, as the midplane will be colder than 40~K outside of $R\gtrsim5$~AU.  A planet in a Hill-radius sized gap will heat the gap edges by $\sim10-20$~K, which will sublimate certain species and greatly enhance the column density of these species locally.  These changes in column density will translate to observable signatures in the circumstellar disk emission, which will appear as emission line asymmetries.  Such asymmetries can be highlighted by subtracting off the azimuthally averaged line brightness profile of the circumstellar disk.  We found that HCN isotopologues are a particularly robust tracer of the heating effects of the planet.  Additionally, multiple lines of HCN should be detectable, which will provide additional physical information (temperature) about the planet's local environment. In addition to the excitation information afforded by molecular line observations, the use of gas tracers (as opposed to continuum) will also provide spectral information that can be used to look for velocity substructure near the planet. This technique is most useful for planets between $R\sim10-30$~AU, where the inner disk is innately too warm to provide sufficient temperature contrast, and planets in the outer disk are expected to carve larger gaps (have larger $R_{\rm Hill}$), and as a direct consequence the planet's heating of the disk is more diluted at larger radii.

Finally, the primary (sublimated) species can trigger additional chemical processing, creating so-called ``secondary'' planet tracers.  These secondary products may be even brighter than the primary products;  however, to fully characterize them we must take into account the differential rotation of the disk, which is beyond the scope of the present work, but will be the subject of a follow-up paper.  If the timescales for secondary product formation are a substantial fraction of the orbital time, the local effects may be sheared out over a large azimuthal range, i.e., as an ``arc.''  If the relevant chemical timescales are much longer than the orbital time, the emission signature of the planet will become a ring (or double ring) at the gap edges. However, in the simple case of rapid sublimation/condensation, the chemical signature is predicted to ``follow'' the planet, implying that we will be able to re-observe and confirm that the emission is associated with the planet as it traverses its orbit.

\acknowledgements{Acknowledgements: The authors are grateful for comments from an anonymous referee, which have improved this manuscript.  The authors also thank Bruce Macintosh, Neal Turner and Ewine van Dishoeck for useful discussions.   LIC and EAB acknowledge the support of NSF grant AST-1008800 and the Rackham Predoctoral Fellowship. TJH acknowledges funding from STFC grant ST/J001627/1. Some calculations for this paper were performed on the University of Exeter Supercomputer, a DiRAC Facility jointly funded by STFC, the Large Facilities Capital Fund of BIS, and the University of Exeter.}


\begin{thebibliography}{52}
\expandafter\ifx\csname natexlab\endcsname\relax\def\natexlab#1{#1}\fi

\bibitem[{{Ayliffe} \& {Bate}(2009)}]{ayliffe2009}
{Ayliffe}, B.~A. \& {Bate}, M.~R. 2009, \mnras, 397, 657

\bibitem[{{Bergin} {et~al.}(2014){Bergin}, {Cleeves}, {Crockett}, \&
  {Blake}}]{bergin2014}
{Bergin}, E.~A., {Cleeves}, L.~I., {Crockett}, N., \& {Blake}, G.~A. 2014,
  Faraday Discussions, 168, 61

\bibitem[{{Bethell} \& {Bergin}(2011{\natexlab{a}})}]{bethell2011xr}
{Bethell}, T.~J. \& {Bergin}, E.~A. 2011{\natexlab{a}}, \apj, 740, 7

\bibitem[{{Bethell} \& {Bergin}(2011{\natexlab{b}})}]{bethell2011code}
---. 2011{\natexlab{b}}, \apj, 739, 78

\bibitem[{{Brinch} \& {Hogerheijde}(2010)}]{brinch2010}
{Brinch}, C. \& {Hogerheijde}, M.~R. 2010, \aap, 523, A25

\bibitem[{Bryden {et~al.}(1999)Bryden, Chen, Lin, Nelson, \&
  Papaloizou}]{bryden1999}
Bryden, G., Chen, X., Lin, D. N.~C., Nelson, R.~P., \& Papaloizou, J. C.~B.
  1999, The Astrophysical Journal, 514, 344

\bibitem[{{Cleeves} {et~al.}(2013){Cleeves}, {Adams}, \&
  {Bergin}}]{cleeves2013a}
{Cleeves}, L.~I., {Adams}, F.~C., \& {Bergin}, E.~A. 2013, \apj, 772, 5

\bibitem[{{Cleeves} {et~al.}(2014{\natexlab{a}}){Cleeves}, {Bergin}, \&
  {Adams}}]{cleeves2014par}
{Cleeves}, L.~I., {Bergin}, E.~A., \& {Adams}, F.~C. 2014{\natexlab{a}}, \apj,
  794, 123

\bibitem[{{Cleeves} {et~al.}(2014{\natexlab{b}}){Cleeves}, {Bergin},
  {Alexander}, {Du}, {Graninger}, {{\"O}berg}, \& {Harries}}]{cleeves2014wat}
{Cleeves}, L.~I., {Bergin}, E.~A., {Alexander}, C.~M.~O., {Du}, F.,
  {Graninger}, D., {{\"O}berg}, K.~I., \& {Harries}, T.~J. 2014{\natexlab{b}},
  Science, 345, 1590

\bibitem[{{Cleeves} {et~al.}(2011){Cleeves}, {Bergin}, {Bethell}, {Calvet},
  {Fogel}, {Sauter}, \& {Wolf}}]{cleeves2011}
{Cleeves}, L.~I., {Bergin}, E.~A., {Bethell}, T.~J., {Calvet}, N., {Fogel},
  J.~K.~J., {Sauter}, J., \& {Wolf}, S. 2011, \apjl, 743, L2

\bibitem[{{Cleeves} {et~al.}(2015){Cleeves}, {Bergin}, {Qi}, {Adams}, \&
  {{\"O}berg}}]{cleeves2015tw}
{Cleeves}, L.~I., {Bergin}, E.~A., {Qi}, C., {Adams}, F.~C., \& {{\"O}berg},
  K.~I. 2015, \apj, 799, 204

\bibitem[{{Collings} {et~al.}(2004){Collings}, {Anderson}, {Chen}, {Dever},
  {Viti}, {Williams}, \& {McCoustra}}]{collings2004}
{Collings}, M.~P., {Anderson}, M.~A., {Chen}, R., {Dever}, J.~W., {Viti}, S.,
  {Williams}, D.~A., \& {McCoustra}, M.~R.~S. 2004, \mnras, 354, 1133

\bibitem[{Collings {et~al.}(2003)Collings, Dever, Fraser, McCoustra, \&
  Williams}]{collings2003}
Collings, M.~P., Dever, J.~W., Fraser, H.~J., McCoustra, M. R.~S., \& Williams,
  D.~A. 2003, The Astrophysical Journal, 583, 1058

\bibitem[{{Dong} {et~al.}(2014){Dong}, {Zhu}, \& {Whitney}}]{dong2014}
{Dong}, R., {Zhu}, Z., \& {Whitney}, B. 2014, ArXiv e-prints

\bibitem[{{Draine} \& {Lee}(1984)}]{draine1984}
{Draine}, B.~T. \& {Lee}, H.~M. 1984, \apj, 285, 89

\bibitem[{{Fogel} {et~al.}(2011){Fogel}, {Bethell}, {Bergin}, {Calvet}, \&
  {Semenov}}]{fogel2011}
{Fogel}, J.~K.~J., {Bethell}, T.~J., {Bergin}, E.~A., {Calvet}, N., \&
  {Semenov}, D. 2011, \apj, 726, 29

\bibitem[{Furlan {et~al.}(2006)Furlan, Hartmann, Calvet, D'Alessio,
  Franco-Hern{\'a}ndez, Forrest, Watson, Uchida, Sargent, Green, Keller, \&
  Herter}]{furlan2006}
Furlan, E., Hartmann, L., Calvet, N., D'Alessio, P., Franco-Hern{\'a}ndez, R.,
  Forrest, W.~J., Watson, D.~M., Uchida, K.~I., Sargent, B., Green, J.~D.,
  Keller, L.~D., \& Herter, T.~L. 2006, The Astrophysical Journal Supplement
  Series, 165, 568

\bibitem[{{Furuya} \& {Aikawa}(2014)}]{furuya2014}
{Furuya}, K. \& {Aikawa}, Y. 2014, \apj, 790, 97

\bibitem[{{Garrod} \& {Herbst}(2006)}]{garrod2006}
{Garrod}, R.~T. \& {Herbst}, E. 2006, \aap, 457, 927

\bibitem[{{Harries}(2000)}]{harries2000}
{Harries}, T.~J. 2000, \mnras, 315, 722

\bibitem[{{Harries} {et~al.}(2004){Harries}, {Monnier}, {Symington}, \&
  {Kurosawa}}]{harries2004}
{Harries}, T.~J., {Monnier}, J.~D., {Symington}, N.~H., \& {Kurosawa}, R. 2004,
  \mnras, 350, 565

\bibitem[{{Hartmann} \& {Kenyon}(1985)}]{hartmann1985}
{Hartmann}, L. \& {Kenyon}, S.~J. 1985, \apj, 299, 462

\bibitem[{{Hasegawa} \& {Herbst}(1993)}]{hasegawa1993}
{Hasegawa}, T.~I. \& {Herbst}, E. 1993, \mnras, 261, 83

\bibitem[{{Hasegawa} {et~al.}(1992){Hasegawa}, {Herbst}, \& {Leung}}]{hhl1992}
{Hasegawa}, T.~I., {Herbst}, E., \& {Leung}, C.~M. 1992, \apjs, 82, 167

\bibitem[{{Heays} {et~al.}(2014){Heays}, {Visser}, {Gredel}, {Ubachs}, {Lewis},
  {Gibson}, \& {van Dishoeck}}]{heays2014}
{Heays}, A.~N., {Visser}, R., {Gredel}, R., {Ubachs}, W., {Lewis}, B.~R.,
  {Gibson}, S.~T., \& {van Dishoeck}, E.~F. 2014, \aap, 562, A61

\bibitem[{{Jang-Condell}(2009)}]{jangcondell2009}
{Jang-Condell}, H. 2009, \apj, 700, 820

\bibitem[{{Jang-Condell} \& {Turner}(2012)}]{jangcondell2012}
{Jang-Condell}, H. \& {Turner}, N.~J. 2012, \apj, 749, 153

\bibitem[{{Kley} \& {Nelson}(2012)}]{kley2012}
{Kley}, W. \& {Nelson}, R.~P. 2012, \araa, 50, 211

\bibitem[{{Kraus} \& {Ireland}(2012)}]{kraus2012}
{Kraus}, A.~L. \& {Ireland}, M.~J. 2012, \apj, 745, 5

\bibitem[{{Kurosawa} {et~al.}(2004){Kurosawa}, {Harries}, {Bate}, \&
  {Symington}}]{kurosawa2004}
{Kurosawa}, R., {Harries}, T.~J., {Bate}, M.~R., \& {Symington}, N.~H. 2004,
  \mnras, 351, 1134

\bibitem[{{Lin} \& {Papaloizou}(1986)}]{lin1986}
{Lin}, D.~N.~C. \& {Papaloizou}, J. 1986, \apj, 307, 395

\bibitem[{{Lin} \& {Papaloizou}(1993)}]{lin1993}
{Lin}, D.~N.~C. \& {Papaloizou}, J.~C.~B. 1993, in Protostars and Planets III,
  ed. E.~H. {Levy} \& J.~I. {Lunine}, 749--835

\bibitem[{{Lubow} \& {Martin}(2012)}]{lubow2012}
{Lubow}, S.~H. \& {Martin}, R.~G. 2012, \apjl, 749, L37

\bibitem[{{Lubow} {et~al.}(1999){Lubow}, {Seibert}, \&
  {Artymowicz}}]{lubow1999}
{Lubow}, S.~H., {Seibert}, M., \& {Artymowicz}, P. 1999, \apj, 526, 1001

\bibitem[{{Lucy}(1999)}]{lucy}
{Lucy}, L.~B. 1999, \aap, 344, 282

\bibitem[{{Madhusudhan}(2012)}]{madhu2012}
{Madhusudhan}, N. 2012, \apj, 758, 36

\bibitem[{{Mathis} {et~al.}(1977){Mathis}, {Rumpl}, \& {Nordsieck}}]{mrn1977}
{Mathis}, J.~S., {Rumpl}, W., \& {Nordsieck}, K.~H. 1977, \apj, 217, 425

\bibitem[{{Montesinos} {et~al.}(2015){Montesinos}, {Cuadra}, {Perez},
  {Baruteau}, \& {Casassus}}]{montesinos2015}
{Montesinos}, M., {Cuadra}, J., {Perez}, S., {Baruteau}, C., \& {Casassus}, S.
  2015, ArXiv e-prints

\bibitem[{{{\"O}berg} {et~al.}(2011){{\"O}berg}, {Murray-Clay}, \&
  {Bergin}}]{oberg2011co}
{{\"O}berg}, K.~I., {Murray-Clay}, R., \& {Bergin}, E.~A. 2011, \apjl, 743, L16

\bibitem[{{Pinte} {et~al.}(2009){Pinte}, {Harries}, {Min}, {Watson},
  {Dullemond}, {Woitke}, {M{\'e}nard}, \& {Dur{\'a}n-Rojas}}]{pinte2009}
{Pinte}, C., {Harries}, T.~J., {Min}, M., {Watson}, A.~M., {Dullemond}, C.~P.,
  {Woitke}, P., {M{\'e}nard}, F., \& {Dur{\'a}n-Rojas}, M.~C. 2009, \aap, 498,
  967

\bibitem[{{Pringle}(1981)}]{pringle1981}
{Pringle}, J.~E. 1981, \araa, 19, 137

\bibitem[{Sandford \& Allamandola(1990)}]{Sandford:1990fz}
Sandford, S.~A. \& Allamandola, L.~J. 1990, Icarus, 87, 188

\bibitem[{{Sandford} {et~al.}(1988){Sandford}, {Allamandola}, {Tielens}, \&
  {Valero}}]{sandford1988}
{Sandford}, S.~A., {Allamandola}, L.~J., {Tielens}, A.~G.~G.~M., \& {Valero},
  G.~J. 1988, \apj, 329, 498

\bibitem[{{Smith} {et~al.}(2004){Smith}, {Herbst}, \& {Chang}}]{smith2004}
{Smith}, I.~W.~M., {Herbst}, E., \& {Chang}, Q. 2004, \mnras, 350, 323

\bibitem[{{Teske} {et~al.}(2013){Teske}, {Cunha}, {Schuler}, {Griffith}, \&
  {Smith}}]{teske2013}
{Teske}, J.~K., {Cunha}, K., {Schuler}, S.~C., {Griffith}, C.~A., \& {Smith},
  V.~V. 2013, \apj, 778, 132

\bibitem[{{van der Marel} {et~al.}(2013){van der Marel}, {van Dishoeck},
  {Bruderer}, {Birnstiel}, {Pinilla}, {Dullemond}, {van Kempen}, {Schmalzl},
  {Brown}, {Herczeg}, {Mathews}, \& {Geers}}]{vandermarel2013}
{van der Marel}, N., {van Dishoeck}, E.~F., {Bruderer}, S., {Birnstiel}, T.,
  {Pinilla}, P., {Dullemond}, C.~P., {van Kempen}, T.~A., {Schmalzl}, M.,
  {Brown}, J.~M., {Herczeg}, G.~J., {Mathews}, G.~S., \& {Geers}, V. 2013,
  Science, 340, 1199

\bibitem[{{van der Tak} {et~al.}(2007){van der Tak}, {Black}, {Sch{\"o}ier},
  {Jansen}, \& {van Dishoeck}}]{vandertak2007}
{van der Tak}, F.~F.~S., {Black}, J.~H., {Sch{\"o}ier}, F.~L., {Jansen}, D.~J.,
  \& {van Dishoeck}, E.~F. 2007, \aap, 468, 627

\bibitem[{{Viti} {et~al.}(2004){Viti}, {Collings}, {Dever}, {McCoustra}, \&
  {Williams}}]{viti2004}
{Viti}, S., {Collings}, M.~P., {Dever}, J.~W., {McCoustra}, M.~R.~S., \&
  {Williams}, D.~A. 2004, \mnras, 354, 1141

\bibitem[{{Woitke}(1999)}]{woitke1999}
{Woitke}, P. 1999, in Astronomy with Radioactivities, ed. R.~{Diehl} \&
  D.~{Hartmann}, 163

\bibitem[{{Wolf} \& {D'Angelo}(2005)}]{wolf2005}
{Wolf}, S. \& {D'Angelo}, G. 2005, \apj, 619, 1114

\bibitem[{{Zhu}(2015)}]{zhu2015}
{Zhu}, Z. 2015, \apj, 799, 16

\bibitem[{{Zhu} {et~al.}(2009){Zhu}, {Hartmann}, {Gammie}, \&
  {McKinney}}]{zhu2009}
{Zhu}, Z., {Hartmann}, L., {Gammie}, C., \& {McKinney}, J.~C. 2009, \apj, 701,
  620

\end{thebibliography}
\end{document}